\documentclass[aps, prd, twocolumn, amsmath, floats,floatfix, superscriptaddress, nofootinbib,
showpacs]{revtex4}

\usepackage{amssymb}
\usepackage{amsmath}
\usepackage{verbatim}
\usepackage{mathrsfs}
\usepackage{amsfonts}
\usepackage{latexsym}
\usepackage{epsfig}
\usepackage{color}
\usepackage{graphicx,subfigure}
\usepackage{units}
\usepackage{slashbox}
\usepackage{natbib}

\begin{document}

\title{Dynamical formation \\ of a Reissner-Nordstr\"om black hole with scalar hair in a cavity} 

\author{Nicolas Sanchis-Gual}
\affiliation{Departamento de
  Astronom\'{\i}a y Astrof\'{\i}sica, Universitat de Val\`encia,
  Dr. Moliner 50, 46100, Burjassot (Val\`encia), Spain}

\author{Juan Carlos Degollado} 
\affiliation{
Instituto de Ciencias F\'isicas, Universidad Nacional Aut\'onoma de M\'exico,
Apdo. Postal 48-3, 62251, Cuernavaca, Morelos, M\'exico.}

  \author{Carlos Herdeiro}
\affiliation{Departamento de F\'{\i}sica da Universidade de Aveiro and CIDMA, Campus de Santiago, 
3810-183 Aveiro, Portugal}

\author{Jos\'e A. Font}
\affiliation{Departamento de
  Astronom\'{\i}a y Astrof\'{\i}sica, Universitat de Val\`encia,
  Dr. Moliner 50, 46100, Burjassot (Val\`encia), Spain}
\affiliation{Observatori Astron\`omic, Universitat de Val\`encia, C/ Catedr\'atico 
  Jos\'e Beltr\'an 2, 46980, Paterna (Val\`encia), Spain}
  
\author{Pedro J. Montero} 
\affiliation{Max-Planck-Institut f{\"u}r Astrophysik, Karl-Schwarzschild-Str. 1, 85748, Garching 
bei M{\"u}nchen, Germany}


\date{July 2016}


\begin{abstract} 
In a recent letter~\cite{Sanchis-Gual:2015lje}, we presented numerical relativity simulations, solving the full Einstein--Maxwell--Klein-Gordon equations,  of superradiantly unstable Reissner-Nordstr\"om black holes (BHs), enclosed in a cavity. Low frequency, spherical perturbations of a charged scalar field, trigger this instability. The system's evolution was followed into the non-linear regime, until it relaxed into an equilibrium configuration, found to be a
\textit{hairy} BH: a charged horizon in equilibrium with a scalar field condensate, whose phase is oscillating at the (final) critical frequency. 
Here, we investigate 
the impact of adding self-interactions to the scalar field. In particular, we find sufficiently large self-interactions suppress the exponential growth phase, known from linear theory, and promote a non-monotonic behaviour of the scalar field energy. Furthermore, we discuss in detail 
the influence of the various parameters in this model: the initial BH charge, the initial scalar 
perturbation, the scalar field charge, mass,  and the position of the cavity's boundary (mirror). 
We also investigate the ``explosive" non-linear regime previously reported to be akin to a bosenova. 
A mode analysis shows that the ``explosions" can be interpreted as the decay into the BH of modes that exit the superradiant regime.
\end{abstract}


\pacs{
95.30.Sf  
04.70.Bw 
04.25.dg 
}


\maketitle

\section{Introduction}\label{sec:introduction}
In an attempt to summarize the astonishing simplicity of electrovacuum black holes (BHs), John Wheeler famously coined the 
\textit{dictum}: ``black holes have no hair"~\cite{Ruffini:1971bza}. This catchy statement is, obviously, vague and needs 
to be contextualized.  In fact, it is useful to introduce the following terminology, that clearly separates two different 
interpretations of Wheeler's statement.

The \textit{strong no-hair hypothesis}, on the one hand, asserts that stationary, regular (on and outside a horizon), BH 
solutions described by parameters other than ``charges" associated to Gauss laws, \textit{do not exist}.  This is a commonly 
found interpretation in the current literature. Unfortunately for the worshipers of such enormous simplicity, decades of 
research considering different matter fields showed that BHs can indeed have hair -- see~\cite{Herdeiro:2015waa,Volkov:2016ehx} 
for recent reviews. The strong no-hair hypothesis has been falsified, even if one requires physically reasonable matter 
(obeying all energy conditions), and asymptotically flat spacetimes.

The \textit{weak no-hair hypothesis}, on the other hand, demands only that stationary, regular (on and outside a horizon), 
BH solutions described by parameters other than ``charges" associated to Gauss laws, \textit{cannot form dynamically}. 
This is certainly what the proponents of the no-hair hypothesis had in mind (in the context of astrophysics and asymptotically 
flat spacetimes). The status of this version of the hypothesis is less definite. In particular, in asymptotically flat spacetimes 
and to the best of our knowledge, no stationary ``hairy" BH solution has been shown to form dynamically. Indeed, often, 
but not always, the stationary solutions that have been found as counter-examples to the strong no-hair hypothesis are unstable 
against perturbations, and hence unlikely to form dynamically (see an early discussion of this version of the conjecture in~\cite{Bizon:1994dh}).

An interesting new angle concerning the weak no-hair hypothesis arises in the context of an instability of the paradigmatic BH solution of vacuum General Relativity -- the Kerr solution~\cite{Kerr:1963ud} --, which is triggered by fields that can, potentially, form BH hair. 

Bosonic fields scattering off Kerr BHs can extract energy through the classical 
process of \textit{superradiance}~\cite{Brito:2015oca}. For concreteness, let us focus on a scalar field. This occurs when such a field, oscillating with frequency $\omega$ and with an azimuthal quantum number 
$m$, fulfills the condition $\omega$ $<$ $m\Omega_{H}$ 
\cite{Bardeen:1972fi,Starobinsky:1973a,Press:1972zz,Zouros:1979iw,Cardoso:2004nk,Dolan:2007mj}, 
where $\Omega_{H}$ is the horizon angular velocity. By introducing a mass 
term for the scalar field, or a mirror-like boundary condition, superradiant modes can become trapped, ``mining" energy from the BH and growing exponentially in time, thus triggering an instability of the combined BH-scalar field system. Consequently, in this setup, the \textit{bald} Kerr BH is unstable and the scalar field (which is not associated to a Gauss law) grows in time outside the BH. This growth could, in principle, approach an equilibrium configuration, in which the BH becomes hairy, because stationary solutions describing Kerr BHs with (this type of) scalar hair have been recently discovered~\cite{Herdeiro:2014goa,Herdeiro:2015gia,Herdeiro:2015tia}. So, is the endpoint of the superradiant instability, triggered by a massive scalar field a hairy Kerr BH? In other words, does a stationary, asymptotically flat hairy BH form dynamically in this setup, thus falsifying the weak no-hair hypothesis?

While the initial growth of the superradiant
instability described in the previous paragraph can be captured at the linear level, a fully nonlinear approach is required to address its saturation 
and endpoint. This is, however, a remarkably challenging undertaking with current numerical relativity (NR) technology~\cite{Okawa:2014nda,East:2014prd}. Linear analysis studies for 
Kerr BHs \cite{Cardoso:2004nk,Dolan:2013} have shown that the maximum growth rate of the instability 
is so small that it may remain indistinguishable from numerical errors when performing nonlinear 
numerical simulations \cite{Okawa:2014nda}. Whereas the first nonlinear simulations of superradiant 
scattering of gravitational waves off nearly extremal Kerr BHs have been recently carried out~\cite{East:2014prd}, following the evolution of the superradiant instability presents another level of difficulty.

In view of the difficulties just described, is there a technically simpler model that presents similar features to 
the superradiant instability of the Kerr BH in the presence of massive bosonic fields? Indeed, an analogous, but 
technically simpler setup exists. A superradiant instability appears in the case of a charged $i.e.$  
Reissner-Nordstr\"om (RN), BH. In this case, superradiance occurs when a charged scalar field with 
frequency $\omega$ and charge $q$, scattering off a charged BH with charge $Q$ and horizon
electric potential $\phi_{H}$, obeys the condition, $\omega$ $<$ 
$\omega_{c}\,\equiv\,q\phi_{H}$ \cite{Bekenstein:1973ur}. Unlike the Kerr case, 
mirror-like boundary conditions are necessary to trigger superradiance in the RN BH, $i.e.$, a mass term is not sufficient~\cite{Hod:2012zz,Hod:2013eea} (or necessary). Studies 
in the linearized regime have shown that the growth timescale of unstable modes in the RN case is 
significantly shorter than for the Kerr BH and that those unstable modes may be spherically symmetric 
\cite{Herdeiro:2013pia,Hod:2013fvl,Degollado:2013eqa,Degollado:2013bha}. These features suggest taking charge as a surrogate for rotation, and study the non-linear growth of the superradiant instability in the RN BH in a cavity, 
sometimes dubbed \textit{charged BH bomb}.

In a recent \textit{Letter} \cite{Sanchis-Gual:2015lje}, we reported NR simulations, using the full Einstein equations, of the charged BH bomb.  We found that, indeed, the generic final state is a \textit{hairy} BH: a charged horizon, surrounded by a scalar field condensate storing part of the charge and energy of the initial BH. This condensate's phase oscillates at the threshold frequency of the superradiant instability, thus realizing dynamically charged hairy BHs analogous to Kerr BHs with scalar hair~\cite{Herdeiro:2014goa,Herdeiro:2015gia,Herdeiro:2015tia}. The former have been recently constructed as stationary solutions and a subset was shown to be perturbatively stable~\cite{Dolan:2015dha}. Similar results for the superradiantly unstable RN-AdS BH were found in \cite{Bosch:2016vcp}, considering reflecting boundary conditions at the AdS timelike boundary.

The purpose of this paper is to further the investigation of the dynamics of the coupled BH-scalar field system in a cavity, initiated in~\cite{Sanchis-Gual:2015lje}. Whereas our letter provided the generic picture, here we will pay careful attention to the variation of the hair growth process with the different parameters in the setup, 
namely, the BH initial charge, the initial scalar perturbation, the scalar field charge and mass, as well as the radius of the mirror.
Moreover, we consider the effect of adding a (non-linear) self-interaction term to the scalar field. We shall also investigate in more detail the behaviour of the ``explosive" regime, described in~\cite{Sanchis-Gual:2015lje} to be akin to a bosenova, following~\cite{Yoshino:2012kn,Yoshino:2015nsa}. As we shall discuss, a mode analysis renders a simple and clear interpretation of the observed behaviour, confirming the results found in~\cite{Bosch:2016vcp}, for the asymptotically AdS case.  To accomplish these goals, we have performed a number of NR simulations, that will be detailed below. The numerical techniques and the code used are those already described 
in~\cite{Sanchis-Gual:2015bh}.

The paper is organized as follows: In Section~\ref{sec:formalism} 
we present the basic equations and discuss the initial data used in 
our simulations. Section~\ref{sec:numerics} briefly describes our numerical
approach.
In Section~\ref{sec:num_results} we discuss our 
findings and describe some properties of the solutions. Finally, in 
Section~\ref{sec:conclusions} we sum up our concluding remarks. One Appendix describes some technical details. Throughout the paper Greek 
indices run over spacetime indices (0 to 3), while 
Latin indices run over space indices only (1 to 3). We use units in which $c=G=\hbar=4\pi\epsilon_0=1$.

\section{Basic equations}\label{sec:formalism}

We shall investigate the dynamics of a complex scalar field, with charge $q$ and mass $\mu$, around a 
RN BH, by solving numerically  the fully non-linear Einstein-Maxwell-Klein-Gordon (EMKG) equations, 
described by the action $\mathcal{S}=\int d^4x \sqrt{-g}\mathcal{L}$, where the Lagrangian density is
\begin{equation}
\mathcal{L}=\frac{R-F_{\alpha\beta}F^{\alpha\beta}}{16\pi}-\frac{1}{2}D_\alpha \Phi (D^\alpha\Phi)^*-\frac{\mu^2}{2}|\Phi|^2 - V_{\rm{int}} \ ,
\label{model}
\end{equation}
where $V_{\rm{int}}=\frac{1}{4}\,\lambda\,|\Phi|^{4}$ is a quartic self-interaction potential with 
coupling $\lambda$. We have denoted by $R$ the Ricci scalar, $F_{\alpha \beta}\equiv \nabla_{\alpha}A_{\beta} - \nabla_{\beta}A_{\alpha}$, 
$A_{\alpha}$ is the electromagnetic potential, $D_\alpha$ is the gauge covariant derivative, 
$D_\alpha\equiv \nabla_\alpha +iqA_\alpha$, and $q$ and $\mu$ are the charge and the mass of  the scalar field. 

Varying the above action with respect to the metric yields the Einstein equations, 
$G_{\alpha\beta}=8\pi (T^{\rm SF}_{\alpha\beta}+T_{\alpha\beta}^{\rm EM})$, with the following energy-momentum tensors
\begin{eqnarray}
T^{\rm SF}_{\alpha\beta}&=&\frac{1}{2}(D_{\alpha}\Phi )^{*}(D_{\beta}\Phi )+\frac{1}{2}(D_{\alpha}\Phi)(D_{\beta}\Phi )^{*}\nonumber\\
&-&\frac{1}{2}g_{\alpha\beta}(D^{\sigma}\Phi)^{*}(D_{\sigma}\Phi)
-\frac{\mu^2}{2}g_{\alpha\beta}|\Phi \Phi^{*}| \ \nonumber \\
&-&\frac{1}{4}\,\lambda\,g_{\alpha\beta}|\Phi\Phi^{*}|^{2}, \\
T^{\rm EM}_{\alpha\beta}&=&\frac{1}{4\pi}F_{\alpha\sigma}F^{\sigma}_{\beta}-\frac{1}{16\pi}g_{\alpha\beta}F_{\sigma\delta}F^{\sigma\delta} .
\label{eq:tmunu}
\end{eqnarray}
Varying \eqref{model} with respect to the scalar field yields the Klein-Gordon equation:
\begin{eqnarray}
 \nabla^{\alpha}\nabla_{\alpha} \Phi &+& i q A^{\alpha}(2\nabla_{\alpha}\Phi+iqA_{\alpha}\Phi) \nonumber\\
 &+& iq\Phi\nabla_{\alpha} A^{\alpha}-\mu^2\Phi-\lambda|\Phi|^{2}\Phi=0\ .
\label{eq:KG}
\end{eqnarray}
Finally, varying the action with respect to the Maxwell potential yields the Maxwell equations
\begin{equation}
\nabla^\alpha F_{\alpha\beta}=2\pi i q\left[\Phi^*D_{\beta}\Phi-\Phi(D_{\beta}\Phi)^*\right]:= 4\pi (j_{em})_\beta \ .
\end{equation}
We follow the convention that $\Phi$ is dimensionless and $\mu$ has
dimensions of (length)$^{-1}$. 

In the following we present the explicit evolution equations
we solve in our simulations. While we mainly include this information to make the paper self-contained, 
we keep these sections as concise as possible, and refer the interested reader to~\cite{Sanchis-Gual:2015bh} for further details. The equations are presented for the particular case of spherical symmetry.

\subsection{Spacetime and electromagnetic split}
The 3+1 metric split takes the form:
\begin{equation}
ds^2=(-\alpha^2+\beta^r \beta_r)dt^2+2\beta_r dtdr+e^{4\chi}\left[a\, dr^2+ b\, r^2 d\Omega^2\right] \ ,
\end{equation}
where the lapse $\alpha$, shift component $\beta^r$, and the (spatial) metric functions, $\chi,a,b$ depend only on $t$ and $r$. 

We use the following 3+1 decomposition of the vector field $A^{\alpha}$
\begin{eqnarray}
\varphi&:=&-n_{\nu}A^{\nu}\ ,\\
a^{r}&:=&^{(3)}A^{r}=\gamma^{r}_{\mu}A^{\mu}\ ,
\end{eqnarray}
where $n^\mu$ is the 4-velocity of the Eulerian observer~\cite{Torres:2014fga} and $\gamma_{\mu\nu}=g_{\mu\nu}+n^\mu n_\nu$ is the metric on the spatial slices (first fundamental form). This split defines the scalar and vector electromagnetic potentials measured by Eulerian observers. In our spherically symmetric setup, the electric field $E^\mu=F^{\mu\nu} n_\nu$ has only a radial component and the magnetic field $B^\mu=\star F^{\mu\nu} n_\nu$ vanishes. Spherical symmetry implies we only have to consider the equations for the electric potential, $\varphi$, for the 
radial component of the vector potential, $a^{r}$, and for the radial component of the electric field, $E^{r}$. The evolution equations for these fields and the electric field take the form
\begin{eqnarray}
\partial_{t}\varphi&=&\beta^{r}\partial_{r}\varphi+\alpha K \varphi\nonumber\\
&-&\frac{\alpha}{ae^{4\chi}}\biggl[\partial_{r}a_{r} +a_{r}\biggl(\frac{2}{r}-\frac{\partial_{r}a}{2a}+\frac{\partial{r}b}{b}+2\partial_{r}\chi\biggl)\biggl]\nonumber\\
&-&\frac{a_{r}}{ae^{4\chi}}\partial_{r}\alpha\ ,\\
\partial_{t}a_{r}&=&\beta^{r}\partial_{r}a_{r}+a_{r}\partial_{r}\beta^{r}-\alpha a e^{4\chi}E^{r}-\partial_{r}(\alpha\varphi)\ ,\\
\partial_{t}E^{r}&=&\beta^{r}\partial_{r}E^{r}-E^{r}\partial_{r}\beta^{r}+\alpha K E^{r} - 4\alpha\pi j_{e}^{r}\ ,
\end{eqnarray}
where $K$ is the trace of the extrinsic curvature $K_{ij}$ (the second fundamental form) and $j_{e}^{r}$ is the electric current density measured by Eulerian observers.

\subsection{Charged Klein-Gordon equation}

To solve the Klein-Gordon equation we introduce two first-order variables, defined as:
\begin{eqnarray}
\Pi &:=& n^{\alpha}\partial_{\alpha}\Phi=\frac{1}{\alpha}(\partial_{t}\Phi-\beta^{r}\partial_{r}\Phi) \ ,\\
\Psi&:=&\partial_{r}\Phi \ .
\end{eqnarray}
Therefore, using Eq.~(\ref{eq:KG}) we obtain the following system of first-order equations: 
\begin{eqnarray}
\partial_{t}\Phi&=&\beta^{r}\partial_{r}\Phi+\alpha\Pi \ ,\\
\partial_{t}\Psi&=&\beta^{r}\partial_{r}\Psi+\Psi\partial_{r}\beta^{r}+\partial_{r}(\alpha\Pi) \ ,\\
\partial_{t}\Pi&=&\beta^{r}\partial_{r}\Pi+\frac{\alpha}{ae^{4\chi}}\biggl[\partial_{r}\Psi\nonumber\\
&+&\Psi\biggl(\frac{2}{r}-\frac{\partial_{r}a}{2a}+\frac{\partial_{r}b}{b}+2\partial_{r}\chi\biggl)\biggl]\nonumber\\
&-& \alpha\biggl[\mu^{2} +\lambda\,|\Phi|^{2} + q^{2}\biggl(\frac{a_{r}^{2}}{ae^{4\chi}}-\varphi^{2}\biggl)\biggl]\Phi+\alpha K\Pi\nonumber\\
&+&\frac{\Psi}{ae^{4\chi}}\partial_{r}\alpha+ 2iq\alpha\biggl[\frac{a_{r}\Psi}{ae^{4\chi}}+\varphi\Pi\biggl]\ .
\label{eq:sist-KG}
\end{eqnarray}

\subsection{Energy-Momentum tensor}

We define the gauge invariant versions of the variables $\Pi$ and $\Psi$
\begin{eqnarray}
\tilde\Pi&:=&n^{\mu}\mathcal{D}_{\mu}\Phi^{*}=\Pi-iq\varphi\Phi\ ,\\
\tilde\Psi&:=&\gamma^{\mu}_{r}\mathcal{D}_{\mu}\Phi=\Psi+iqa_{r}\Phi\ .
\end{eqnarray}
The matter source terms  for the scalar field read
\begin{eqnarray}
\mathcal{E}^{\rm{SF}}&:=&n^{\alpha}n^{\beta}T^{\rm{SF}}_{\alpha\beta}=\frac{1}{2}\biggl(|\tilde\Pi|^{2}+\frac{|\tilde\Psi|^{2}}{ae^{4\chi}}\biggl) \nonumber\\
&&+\frac{1}{2}\mu^{2}|\Phi|^{2}+\frac{1}{4}\lambda\,|\Phi|^{4} \label{eq:rho}, \label{scalared}\\
j_{r}^{\rm{SF}}&:=&-\gamma^{\alpha}_{r}n^{\beta}T^{\rm{SF}}_{\alpha\beta}=-\frac{1}{2}\biggl(\tilde\Pi^{*}\tilde\Psi+\tilde\Psi^{*}\tilde\Pi\biggl)\ ,\\
S_{a}^{\rm{SF}}&:=&(T^{r}_{r})^{\rm{SF}}=\frac{1}{2}\biggl(|\tilde\Pi|^{2}+\frac{|\tilde\Psi|^{2}}{ae^{4\chi}}\biggl) \nonumber\\
&&-\frac{1}{2}\mu^{2}|\Phi|^{2}-\frac{1}{4}\lambda\,|\Phi|^{4} \ ,\\
S_{b}^{\rm{SF}}&:=&(T^{\theta}_{\theta})^{\rm{SF}}=\frac{1}{2}\biggl(|\tilde\Pi|^{2}-\frac{|\tilde\Psi|^{2}}{ae^{4\chi}}\biggl)  \nonumber\\
&&-\frac{1}{2}\mu^{2}|\Phi|^{2}-\frac{1}{4}\lambda\,|\Phi|^{4}\ .
\end{eqnarray}
and for the electric field
\begin{eqnarray}
\mathcal{E}^{\rm{em}}&=&\frac{1}{8\pi}\,a\,e^{4\chi}(E^{r})^{2}\ ,\\
S_{a}^{\rm{em}}&=&-\frac{1}{8\pi}\,a\,e^{4\chi}(E^{r})^{2}\ ,\\
S_{b}^{\rm{em}}&=&\frac{1}{8\pi}\,a\,e^{4\chi}(E^{r})^{2} \ .
\end{eqnarray}

The momentum density $j_{r}^{\rm{em}}$ vanishes because there is no magnetic field in spherical symmetry.
\subsection{Initial data}

As in our Letter \cite{Sanchis-Gual:2015lje}, we choose the initial data for the scalar field to be a Gaussian distribution, of the form
\begin{equation}\label{eq:pulse}
 \Phi=A_0e^{-(r-r_0)^2/\sigma^2} \ ,
\end{equation}
where $A_0$ is the initial amplitude of the pulse, $r_0$ is the center of the Gaussian, and
$\sigma$ is its width. This scalar field will always be contained within a cavity, whose boundary we call ``the mirror".

The auxiliary first order quantities are
initialized as follows
\begin{eqnarray}
 \Pi(t=0,r)&=&0 \ , \\ 
 \Psi(t=0,r) &=& -2\frac{(r-r_0)}{\sigma^2}A_0e^{-(r-r_0)^2/\sigma^2} \ .
 \label{eq:iderivatives}
\end{eqnarray}

As the geometrical initial data, we choose a conformally flat metric with $a=b=1$ together with a time symmetry condition $K_{ij}=0$. 
This describes a time-slice of a RN BH, in isotropic coordinates, if the 3-metric is written as
\begin{equation}
dl^{2}=\psi^{4}(dr^{2}+r^{2}d\Omega^{2}) \ ,
\end{equation}
and the conformal factor is given by
\begin{equation}
\psi = \biggl[\biggl(1+\frac{M}{2r}\biggl)^{2}-\frac{Q^{2}}{4r^{2}}\biggl]^{1/2},
\end{equation}\\
where $M$ is the BH mass and $Q$ its charge. 

At $t=0$, we choose a ``pre-collapsed" lapse 
\begin{equation}
\alpha = \psi^{-2}\ ,
\end{equation}
and a vanishing shif $\beta^{r}=0$. Initially, the electric field is given by
\begin{equation}
E^{r}=\frac{Q}{r^{2}\psi^{6}} \ .
\end{equation}

The mirror-like boundary conditions are
\begin{eqnarray}\label{eq:mirror}
\Phi(r_{\rm{m}})=\Psi(r_{\rm{m}})=\Pi(r_{\rm{m}})=0\ ,\nonumber \\
\partial_{r}\Phi(r_{\rm{m}})=\partial_{r}\Psi(r_{\rm{m}})=\partial_{r}\Pi(r_{\rm{m}})=0\ .
\end{eqnarray}

\begin{figure}[h!]
\includegraphics[width=0.42\textwidth]{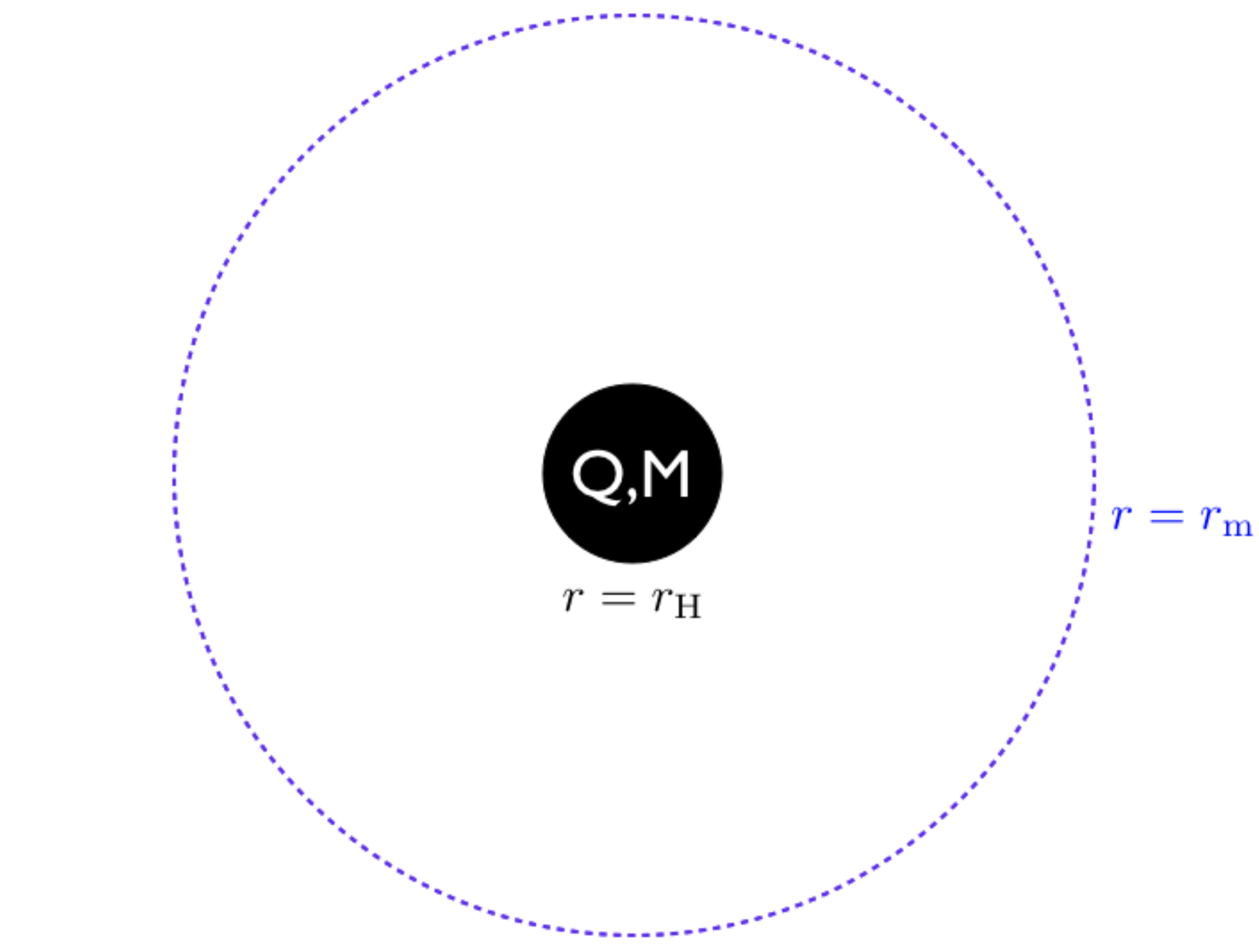} 
\vspace{0.3cm}
\includegraphics[width=0.45\textwidth]{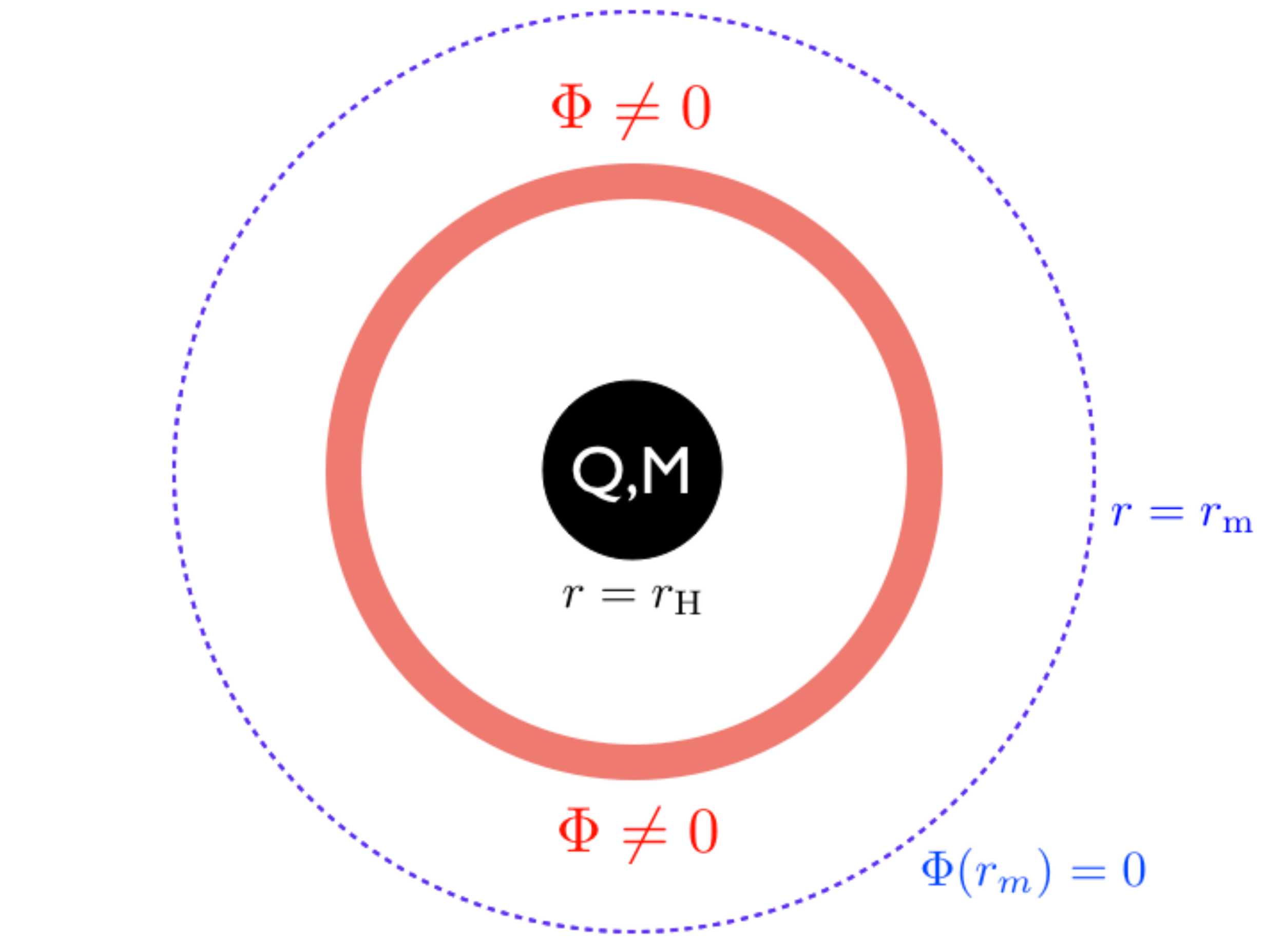} 
\caption{Schematic representation of the RN BH with mass $M$ and charge $Q$ in a cavity with boundary at $r=r_{\rm m}$, where mirror boundary conditions for the scalar field are imposed. Top panel: the unperturbed setup. In this case the cavity is irrelevant since neither the gravitational nor the electromagnetic field have special boundary conditions at the cavity's boundary; Bottom panel: the perturbed setup, setting a Gaussian scalar cloud around the BH. The scalar field obeys reflective boundary conditions at the cavity's boundary (hence called \textit{mirror}).}
\label{fig_unperturbed}
\end{figure}

To summarize, the model (background plus field properties) to be studied contains five parameters:
\begin{equation}
M,Q,r_{\rm m},\mu, q \ .
\end{equation}
In the following we take $M=1$ for all the simulations, which fixes the energy scale of the problem, but will vary the value of $Q$, focusing on the sample 
\begin{equation}
Q=\{0, 0.3, 0.5, 0.7, 0.9\}M \ .
\end{equation} 
The mirror shall be considered at three different positions, with radial coordinates 
\begin{equation}
r_{\rm{m}}=\lbrace9,14.2,19\rbrace M\ ,
\label{rmvalues}
\end{equation}
in order to study its influence in the evolution of the superradiant instability. For the scalar field mass we shall consider both a massless and a massive field:
\begin{equation}
\mu M=\{0, 0.1\} \ , 
\end{equation}
and we consider seven models with different values of the scalar field charge $qM$, namely 
\begin{equation}
qM=\{0.8, 1, 1.2, 2, 5, 10, 20,40\} \ .
\label{qvalues}
\end{equation}
The initial data for the scalar field cloud introduces three other parameters, as described above. 
For all models, except those in Sec.~\ref{sec_varycloud}, we choose $A_0=3\times 10^{-4}$, 
$\sigma=\sqrt{2}$. In Sec.~\ref{sec_varycloud} we also consider
 $A_0 = 2.1\times10^{-5}, \sigma=0.01$ and $A_0 = 2.0\times 10^{-4},\sigma=1.8$.
The center of the Gaussian is $r_0=7M$, when we set the mirror at $r_{\rm{m}}=14.2M$ and $r_{\rm{m}}=19M$; on the other hand,  $r_0=5M$ for $r_{\rm{m}}=9M$. In all simulations below, with the exception of Sec.~\ref{sec_selfint}, we take the self-interaction coupling $\lambda=0$. In Sec.~\ref{sec_selfint} we consider the values
\begin{equation}
\lambda=\{0,1.5, 5.0, 7.5\}\times10^4  \ .
\end{equation}
A schematic representation of the unperturbed and perturbed RN BH in a cavity is exhibited in Fig.~\ref{fig_unperturbed}.

\section{Numerics} 
\label{sec:numerics}

The time update of the different systems of evolution equations we have to solve in our code 
(Einstein, Klein-Gordon, and Maxwell) is done using the same type of techniques we have extensively 
used in previous work (see, in particular,~\cite{Montero:2012yr,Sanchis-Gual:2015bh,Sanchis-Gual:2015sms}). We refer  
the interested reader to those references for full details on the particular numerical techniques 
implemented in the code. Here, we simply mention that the evolution equations are integrated using 
the second-order PIRK method developed by \cite{Isabel:2012arx,Casas:2014}.  This 
method allows to handle the singular terms that appear in the evolution equations due to our choice 
of curvilinear coordinates. The derivatives in the spacetime evolution are computed using a 
fourth-order centered finite difference approximation on a log grid except 
for advection terms for which we adopt a fourth-order upwind scheme.  
We also use fourth-order 
Kreiss-Oliger dissipation to avoid high frequency noise appearing near the outer boundary. In this work 
we are also evolving the electric field explicitly and the electric potentials implicitly.

\section{Results} 
\label{sec:num_results}

\subsection{Initial setup, convergence and constraint violations}

 The EMKG system admits as a solution the RN BH with ADM mass $M$ and charge $Q$, together with a vanishing scalar field. We perturb the RN BH by surrounding it with a charged scalar field cloud whose intial form is given by Eq.~(\ref{eq:pulse}) -- see Fig.~\ref{fig_unperturbed}, bottom panel. The superradiant instability, which leads to the growth of the scalar field outside the horizon, and the loss of energy/charge by the BH, is triggered if the scalar cloud oscillations includes modes with frequency $w<w_{\rm c} \equiv q\phi_H$, where $\phi_H$ is the electric potential at the horizon. The trapping of the superradiant modes, which is fundamental for the instability, is guaranteed by imposing reflecting boundary conditions for the scalar field at the spherical mirror, located at $r=r_{\rm{m}}$. 
 
In the numerical simulations performed to follow the development of the instability, we have used a logarithmic radial grid that extends from the origin to $r=10^4M$ and uses a maximum resolution close to the origin of
$\Delta r=0.025M$. In order to test the convergence of the code we 
performed three simulations with different resolutions $\Delta r = \lbrace0.025,0.0125,0.00625\rbrace$M. In~\cite{Sanchis-Gual:2015lje} (see supplemental material therein) we have already shown the rescaled evolution of the L2 norm of the Hamiltonian constraint 
for a particular choice of the scalar field charge, $qM=40$ and mirror position $r_{\rm{m}}=14.2M$, obtaining the expected second-order convergence of our PIRK time-evolution scheme. We note that the same result is achieved irrespective of the combination of parameters considered.

 We remark that in  our setup,  the  initial  data  do  not  satisfy  the  constraints. Nevertheless, as discussed in detail in the supplemental material in~\cite{Sanchis-Gual:2015lje},  this fact does not introduce significant errors in the simulations.

\subsection{System's evolution: general picture}

\begin{figure}[h!]
\begin{minipage}{1\linewidth}
\includegraphics[width=1.0\textwidth, height=0.3\textheight]{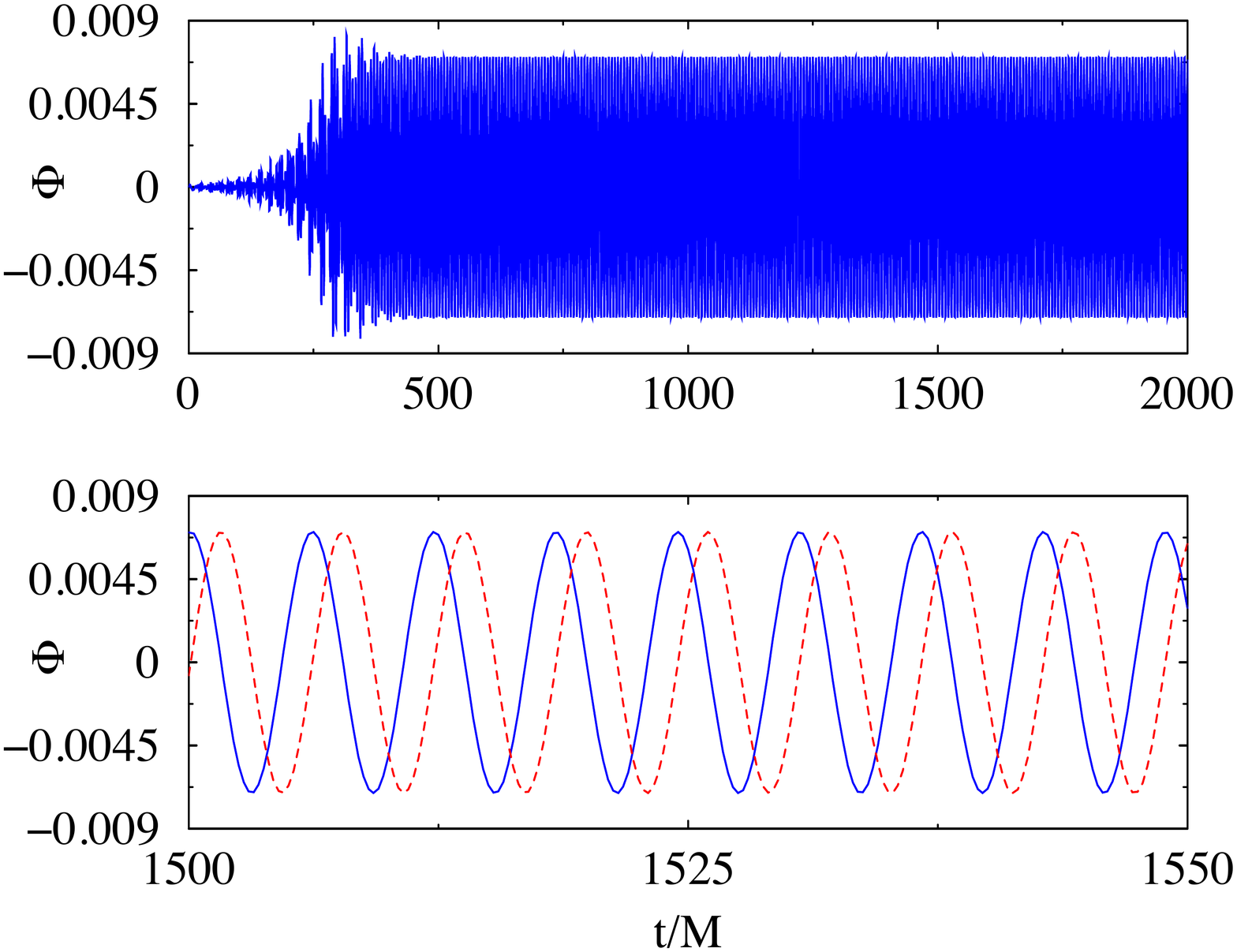} 
\includegraphics[width=1.0\textwidth, height=0.3\textheight]{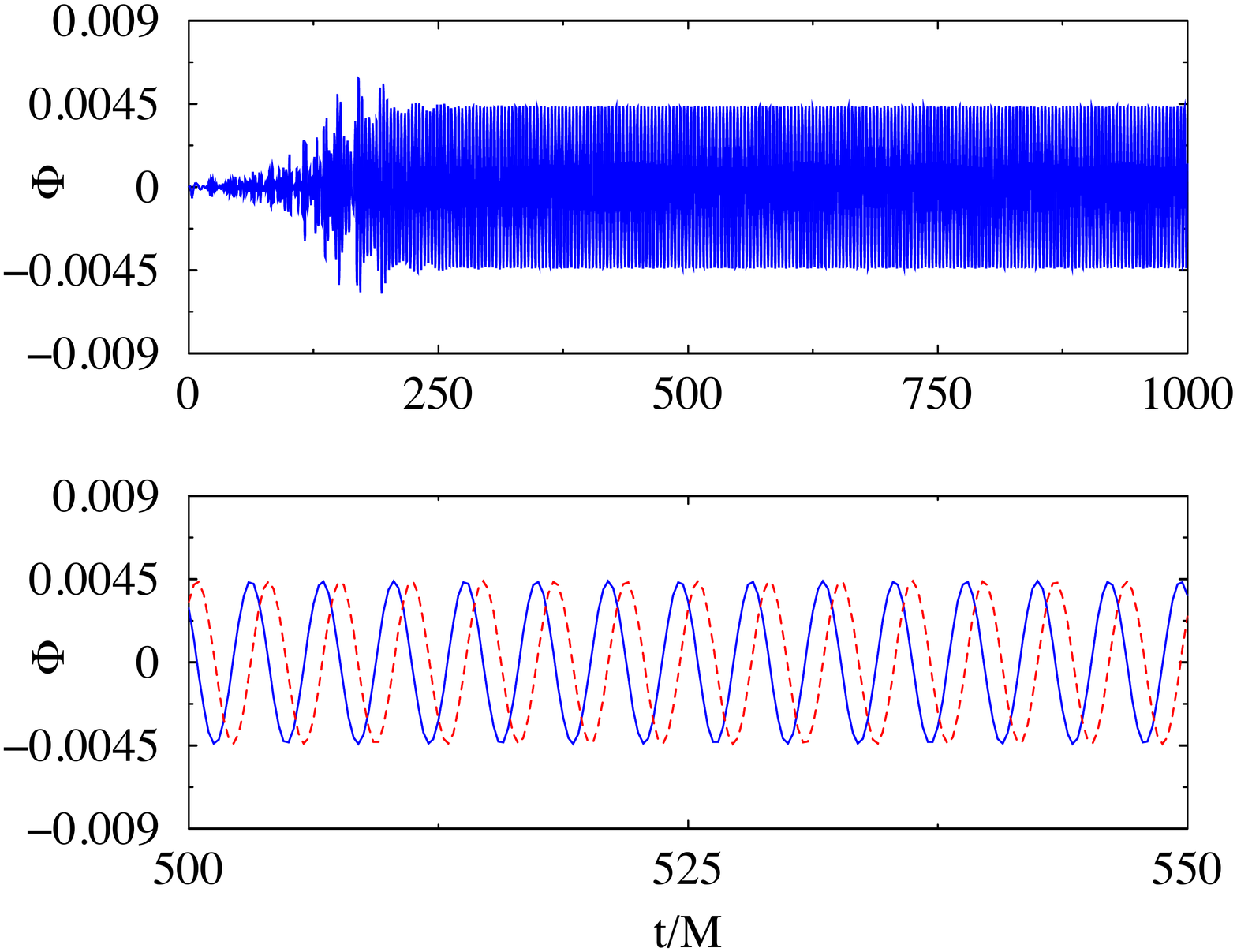} 
\caption{First (top) panel: Time evolution of the scalar field real part, extracted at $r_{\rm{obs}}=5M$, for $qM=10$, $Q=0.9M$, $\mu M=0.1$ and for $r_{\rm{m}}=14.2M$. Second panel: Detail of the time evolution of the 
scalar field real (blue solid line) and imaginary (red dashed line) parts. Third and bottom panels: Same as first and second rows, but for $qM=20$.}
\label{fg:obs1}
\end{minipage}
\end{figure}

We solve numerically the EMKG system using the initial data 
given by Eqs.~(\ref{eq:pulse})-(\ref{eq:mirror}) and let the 
superradiant instability grow. As in \cite{Sanchis-Gual:2015lje} we analyze the results of 
the simulations by extracting a time series for the scalar field 
amplitude at an observation point located at one fixed radii, here taken to be at  
$r_{\rm{obs}}=5M$ (a different value from that used in~\cite{Sanchis-Gual:2015lje}). Typical 
behaviours are shown in Fig. \ref{fg:obs1}. To identify the frequencies at which the scalar 
field oscillates we perform a Fast Fourier transform after a given 
number of time steps and obtain the corresponding power spectrum.

The time evolution of the scalar field amplitude exhibited in Fig.~\ref{fg:obs1} shows two distinct phases. During the first phase -- the \textit{superradiant growth phase} --, the amplitude of the oscillations of the scalar field grow exponentially (at the observation point), which is the expected behaviour due to the superradiant instability, well known from the linear theory analysis~\cite{Herdeiro:2013pia,Hod:2013fvl,Degollado:2013eqa}. During a second phase -- the \textit{saturation and equilibrium phase} --, the exponential growth of the scalar field stops and an equilibrium between the scalar field and the BH is attained, during which the amplitude of the scalar field remains constant and the real and imaginary parts of the scalar field oscillate with a single frequency and with opposite phases ($i.e.$ when one is at a maximum of the magnitude of the amplitude the other one has a vanishing amplitude) -- Fig.~\ref{fg:obs1}, second and fourth rows). These plots show the power spectra obtained from the Fourier transforms of the time series. 

The true nature of this final equilibrium state is revealed by computing also the critical 
frequency $\omega^{\rm fin}_{\rm c}\equiv q\phi_H^{\rm fin}$, from the horizon electric potential of the final BH. The latter is computed at the apparent horizon (AH) of the final BH as~\cite{Alcubierre:2009ij}
\begin{equation}
\phi_H =  
\alpha \varphi-\beta^{r}a_{r}|_{r=r_{\rm{AH}}} \ .
\label{potentialh}
\end{equation}
We obtain \textit{precisely} the same value as that of the final frequency of the scalar field -- see~Table~\ref{tab:mod1}, fourth and fifth columns. The condition $\omega=\omega_{c}$ is thus fulfilled, implying these configurations are \textit{hairy} BHs that exist at the threshold of the superradiant instability. These solutions were first discussed for rotating BHs bifurcating from the Kerr solution in~\cite{Herdeiro:2014goa}, and for charged BHs in a cavity bifurcating from the RN solution in~\cite{Dolan:2015dha}. In particular, the latter paper established that solutions with no nodes in the scalar field profile (like the ones obtained here) are stable against radial perturbations. This provides strong evidence that the equilibrium state obtained herein is the end-point of the evolution.

To summarize: a  RN BH, perturbed by a charged scalar field confined within a cavity around the BH, containing low frequency modes, such that $w<q\Phi_H$, is unstable. During a first phase, the BH transfers part of its energy and charge into the scalar field and the scalar field grows exponentially while the horizon electric potential, $\phi_H$, of the BH decreases. In a second phase this growth stops when a single mode of the scalar field remains, with precisely the critical frequency of the BH,  $q\Phi_H^{\rm final}$. This is the general picture observed in all simulations. Now we shall discuss how this general picture is sensitive to the different parameters of the system.

\subsection{System's evolution: detailed description}

\begin{figure}[h!]
\begin{minipage}{1\linewidth}
\includegraphics[width=1.0\textwidth, height=0.3\textheight]{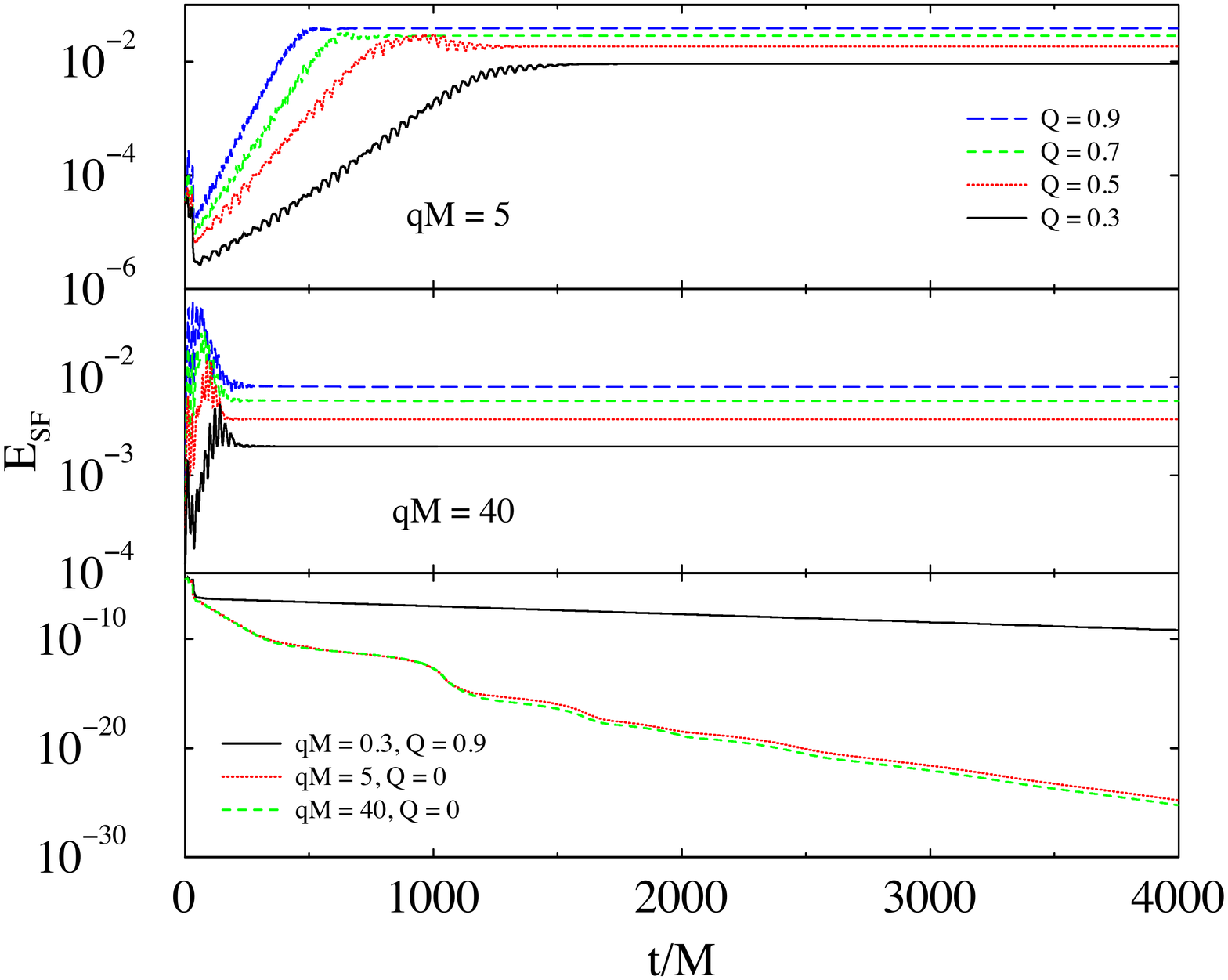} 
\caption{Time evolution of the scalar field energy, displayed in logarithmic scale, for $r_{\rm{m}}=14.2M$, 
different values of the initial BH charge $Q$ and: (top panel), $qM = 5$; (middle panel) $qM = 40$;  (bottom panel) different values of the scalar field charge $q$.}
\label{fg:esf_difQ}
\end{minipage}
\end{figure}

The most relevant dynamics of our system concerns the energy  and charge transfers between the BH and the scalar field. The energy in the scalar field can be computed by the (spatial) volume integral
\begin{equation}
E_{\rm{SF}}=\int^{r_{\rm m}}_{r_{\rm{AH}}}\mathcal{E}^{\rm{SF}}dV\ ,
\label{eq:ESF}
\end{equation}
where $\mathcal{E}^{\rm{SF}}$ is the 
projection of the energy-momentum tensor of the scalar field along the normal direction to the $t= $constant surfaces~\cite{Alcubierre08a}, $cf.$~Eq.~\eqref{scalared}. 
In Fig.~\ref{fg:esf_difQ} we plot the evolution of this scalar field energy for different values of the BH initial charge, $Q$ (and also of the scalar field charge $q$). The first important feature, manifest on the bottom panel, is that for vanishing BH charge the scalar field energy \textit{does not grow}. In other words, there is no superradiant instability of uncharged BHs. The second important trend is that for fixed scalar field charge, the instability is stronger -- both in terms of a shorter time scale as well as in terms of a larger energy transfer into the scalar field -- for larger $Q$ (top and middle panels). Finally, observe that even if both the scalar field and the BH are charged, but if there are no superradiant modes in the scalar field cloud, there is no growth of the scalar field. This is seen in one of the examples in the bottom panel, for which the choice of parameters ($q$ and $Q$), leads to $\omega_{\rm{SF}}>\omega_{\rm{c}}$.

Having clarified the essential trends when varying the BH charge, we fix this charge to a large value $Q=0.9$ to make the superradiant instability strong and focus on the variation of the scalar field charge and the mirror radius. We have evolved $8\times 3=24$ different models to study the variation of these parameters corresponding to the values shown in Eqs.~\eqref{qvalues} and~\eqref{rmvalues}. A summary of the physical quantities obtained in these evolutions is shown in Table I.

\begin{table*}
\caption{Summary of physical quantities for the runs with different values of $qM$ and $r_{\rm{m}} = 9M$ (top table), $r_{\rm{m}}=14.2M$ (middle table) and $r_{\rm{m}}=19M$ (bottom table). Each model (first column) a--h corresponds to the values in equation~\eqref{qvalues}, which are shown in the second column; (third column) e-folding time during the growth phase; (fourth and fifth columns) final oscillation frequency of the scalar field phase and final critical frequency; (sixth to eighth columns) initial and final scalar field energy, and their ratio; (ninth to eleventh columns) final BH irreducible mass and ratio of the final to initial BH and scalar field charge.}
\label{tab:mod1}
\begin{ruledtabular}
\begin{tabular}{c||c|c|cc|ccc|ccc}
Model&$qM$& $\tau/M$&{$\,\,M\omega^{\rm fin}_{\rm SF}$} &$M\omega_{\rm c}^{\rm fin}$&$E_{\rm SF}^{\rm ini}/M$&$E_{\rm SF}^{\rm fin}/M$&$E_{\rm SF}^{\rm fin}/E_{\rm SF}^{\rm ini}$&$M_{\rm{irr}}^{\rm fin}/M$&$Q_{\rm{BH}}^{\rm{fin}}/Q$&$Q_{\rm{SF}}^{\rm{fin}}/Q$\\
\hline
1a&0.8&3.3E02&0.376&0.377&1.66E-05&1.29E-01&7.77E03&0.721&60 \%&40 \%\\
1b&1.0&2.4E02&0.405&0.405&1.67E-05&1.33E-01&7.96E03&0.723&48 \%&52 \%\\
1c&1.2&2.0E02&0.435&0.436&1.69E-05&1.29E-01&7.63E03&0.732&41 \%&59 \%\\
1d&2.0&1.3E02&0.546&0.546&1.81E-05&1.01E-01&5.58E03&0.766&24 \%&76 \%\\
1e&5.0&6.5E01&0.928&0.928&2.79E-05&5.29E-02&1.90E03&0.838&8.0 \% & 92 \% \\
1f&10.0&4.3E01&1.513&1.514&6.27E-05&3.11E-02&4.96E02&0.870&3.0 \%& 97 \%\\
1g&20.0&3.3E01&2.607&2.608&2.02E-04&1.84E-02&9.11E01&0.881&2.0 \%& 98 \%\\
1h&40.0&2.0E01&4.676&4.676&7.59E-04&1.15E-02&1.52E01&0.900&0.6 \% & 99.4 \%\\
\end{tabular}
\end{ruledtabular}
\begin{ruledtabular}
\begin{tabular}{c||c|c|cc|ccc|ccc}
Model&$qM$& $\tau/M$&{$\,\,M\omega^{\rm fin}_{\rm SF}$} &$M\omega_{\rm c}^{\rm fin}$&$E_{\rm SF}^{\rm ini}/M$&$E_{\rm SF}^{\rm fin}/M$&$E_{\rm SF}^{\rm fin}/E_{\rm SF}^{\rm ini}$&$M_{\rm{irr}}^{\rm fin}/M$&$Q_{\rm{BH}}^{\rm{fin}}/Q$&$Q_{\rm{SF}}^{\rm{fin}}/Q$\\
\hline
2a&0.8&4.8E02&0.277&0.278&3.00E-05&1.32E-01&4.40E03&0.728&45 \%&55 \%\\
2b&1.0&3.7E02&0.296&0.297&3.01E-05&1.22E-01&4.05E03&0.742&36 \%&64 \%\\
2c&1.2&3.4E02&0.315&0.316&3.04E-05&1.11E-01&3.65E03&0.764&31 \%&69 \%\\
2d&2.0&2.1E02&0.389&0.390&3.17E-05&8.02E-02&2.53E03&0.815&18 \%&82 \%\\
2e&5.0&1.1E02&0.642&0.642&4.31E-05&3.93E-02&9.12E02&0.875&6.0 \% & 94 \% \\
2f&10.0&7.1E01&1.030&1.031&8.37E-05&2.25E-02&2.69E02&0.903&2.0 \%& 98 \%\\
2g&20.0&4.8E01&1.756&1.756&3.13E-04&1.31E-02&4.19E01&0.924&1.0 \%& 99 \%\\
2h&40.0&2.9E01&3.130&3.129&8.95E-04&8.02E-03&8.96E00&0.942&0.1 \% & 99.9 \%\\
\end{tabular}
\end{ruledtabular}
\begin{ruledtabular}
\begin{tabular}{c||c|c|cc|ccc|ccc}
Model&$qM$& $\tau/M$&{$\,\,M\omega^{\rm fin}_{\rm SF}$} &$M\omega_{\rm c}^{\rm fin}$&$E_{\rm SF}^{\rm ini}/M$&$E_{\rm SF}^{\rm fin}/M$&$E_{\rm SF}^{\rm fin}/E_{\rm SF}^{\rm ini}$&$M_{\rm{irr}}^{\rm fin}/M$&$Q_{\rm{BH}}^{\rm{fin}}/Q$&$Q_{\rm{SF}}^{\rm{fin}}/Q$\\
\hline
3a&0.8&6.3E02&0.231&0.232&2.99E-05&1.19E-01&3.98E03&0.773&40.5 \%&59.5 \%\\
3b&1.0&4.8E02&0.244&0.244&3.01E-05&1.10E-01&3.65E03&0.777&31 \%&69 \%\\
3c&1.2&4.2E02&0.257&0.259&3.04E-05&9.87E-02&3.25E03&0.796&26 \%&74 \%\\
3d&2.0&2.7E02&0.313&0.314&3.17E-05&6.89E-02&2.17E03&0.846&15 \%&85 \%\\
3e&5.0&1.6E02&0.506&0.507&4.31E-05&3.27E-02&7.59E02&0.902&5.0 \% & 95 \% \\
3f&10.0&1.1E02&0.802&0.802&8.37E-05&1.84E-02&2.20E02&0.927&2.0 \%& 98 \%\\
3g&20.0&7.4E01&1.355&1.355&2.46E-04&1.06E-02&4.30E01&0.935&0.9 \%& 99.1 \%\\
3h&40.0&5.0E01&2.402&2.401&8.95E-04&6.41E-03&7.16E00&0.950&0.02 \% & 99.98 \%\\
\end{tabular}
\end{ruledtabular}
\end{table*}

For each model studied, Table I shows: the e-folding time (third column) obtained as the best fit of the form $|\Phi| \sim e^{t/\tau}$ during the growth phase; the final scalar field frequency obtained from a Fast Fourier transform and the final critical frequency, 
obtained from~\eqref{potentialh}; the initial and final scalar field energy, obtained from~\eqref{eq:ESF}, as well as their ratio;   the final BH irreducible mass,  computed in terms of the AH area $A_{\rm{AH}}$, \cite{Christodoulou:1970wf}, on each time slice, as 
\begin{eqnarray}
M_{\rm{AH}} = \sqrt{\frac{A_{\rm{H}}}{16\pi}} \ ;
\end{eqnarray}
and the final scalar field and BH charge, the former being obtained from a formula similar to~(\ref{eq:ESF}) replacing $\mathcal{E}^{\rm{SF}}$ by the charge density, and  the latter, $Q_{\rm{BH}}$, evaluated at the AH as~\cite{Torres:2014fga}
\begin{equation}
Q_{\rm{BH}}=\left(r^{2}e^{6\chi}\sqrt{ab^{2}}E^{r}\right)\big|_{\rm AH} \ .
\end{equation}

In the following subsections we describe various trends that can be observed from the results in the table.

\subsubsection{Entropy growth}

As a first observation we note that, for the initial RN BH, the irreducible mass is $M_{\rm irr}^{\rm ini}\simeq 0.718M$. 
Inspection of the table shows  that the final BH has a larger $M_{\rm irr}$, for all cases. This confirms that the evolution 
abides with the area law and, in this respect, charged superradiance can be regarded as a classical process in BH physics. 
It can also be concluded that, the final irreducible mass grows with the scalar field charge. This is a consequence of two factors: 
$(i)$  scalar fields with a larger charge are more efficient in discharging the BH, transferring its charge to the scalar 
field; $(ii)$ by contrast, the scalar field energy grows less, in terms of the final-to-initial energy ratio, with increasing scalar field charge.

\subsubsection{Impact of the mirror radius and scalar field charge}

\begin{figure}
\begin{minipage}{1\linewidth}
\includegraphics[width=1.0\textwidth, height=0.3\textheight]{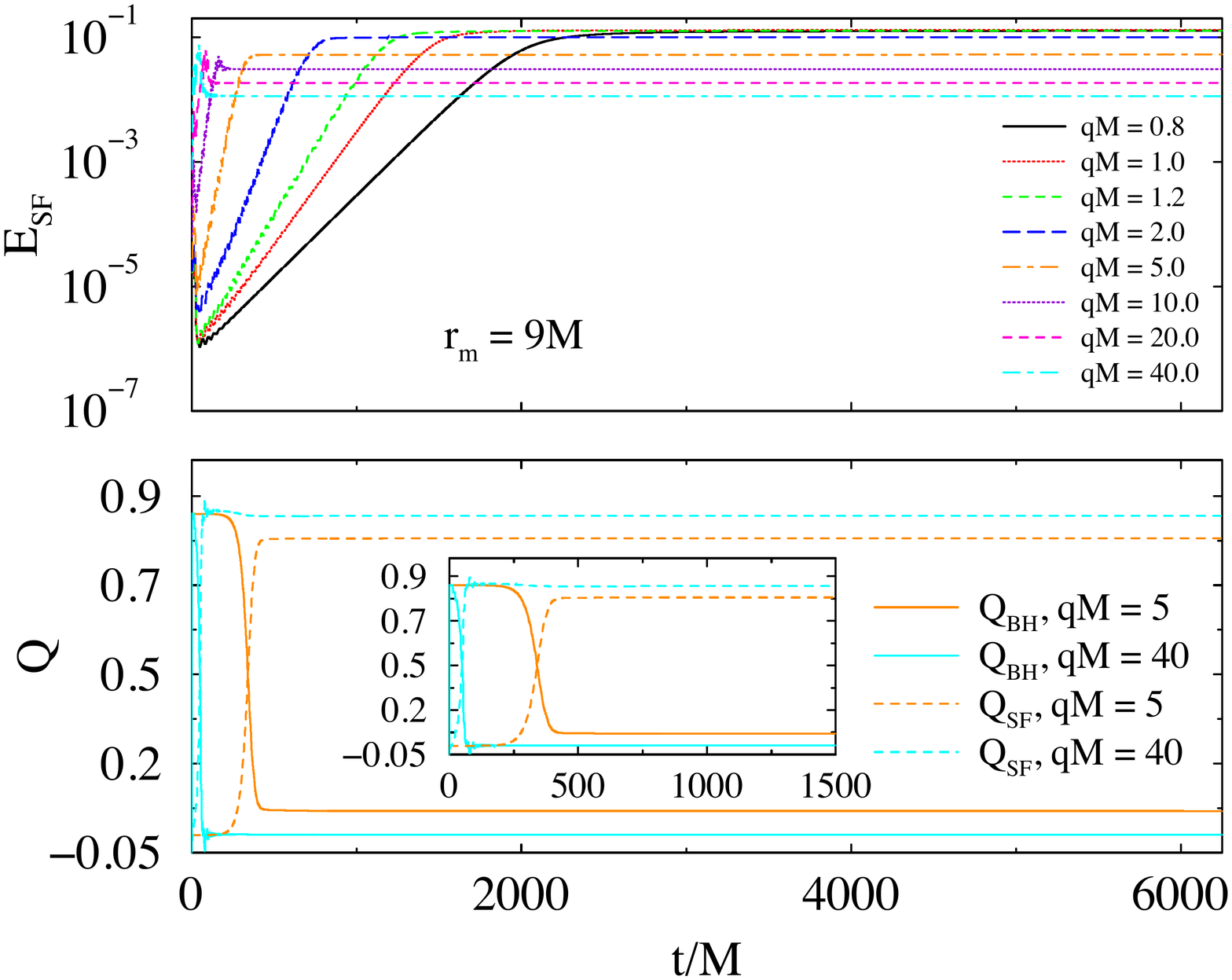} 
\includegraphics[width=1.0\textwidth, height=0.3\textheight]{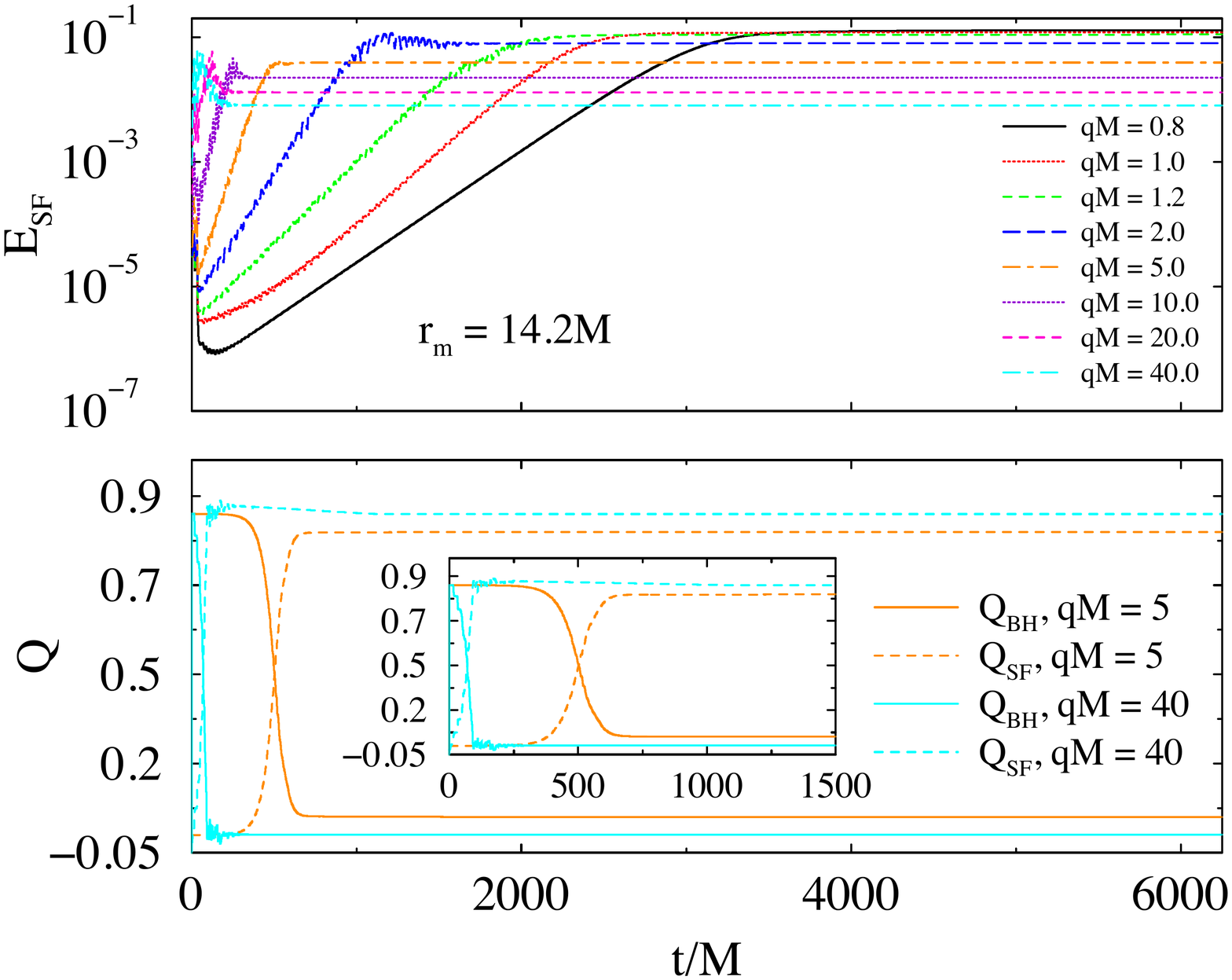} 
\includegraphics[width=1.0\textwidth, height=0.3\textheight]{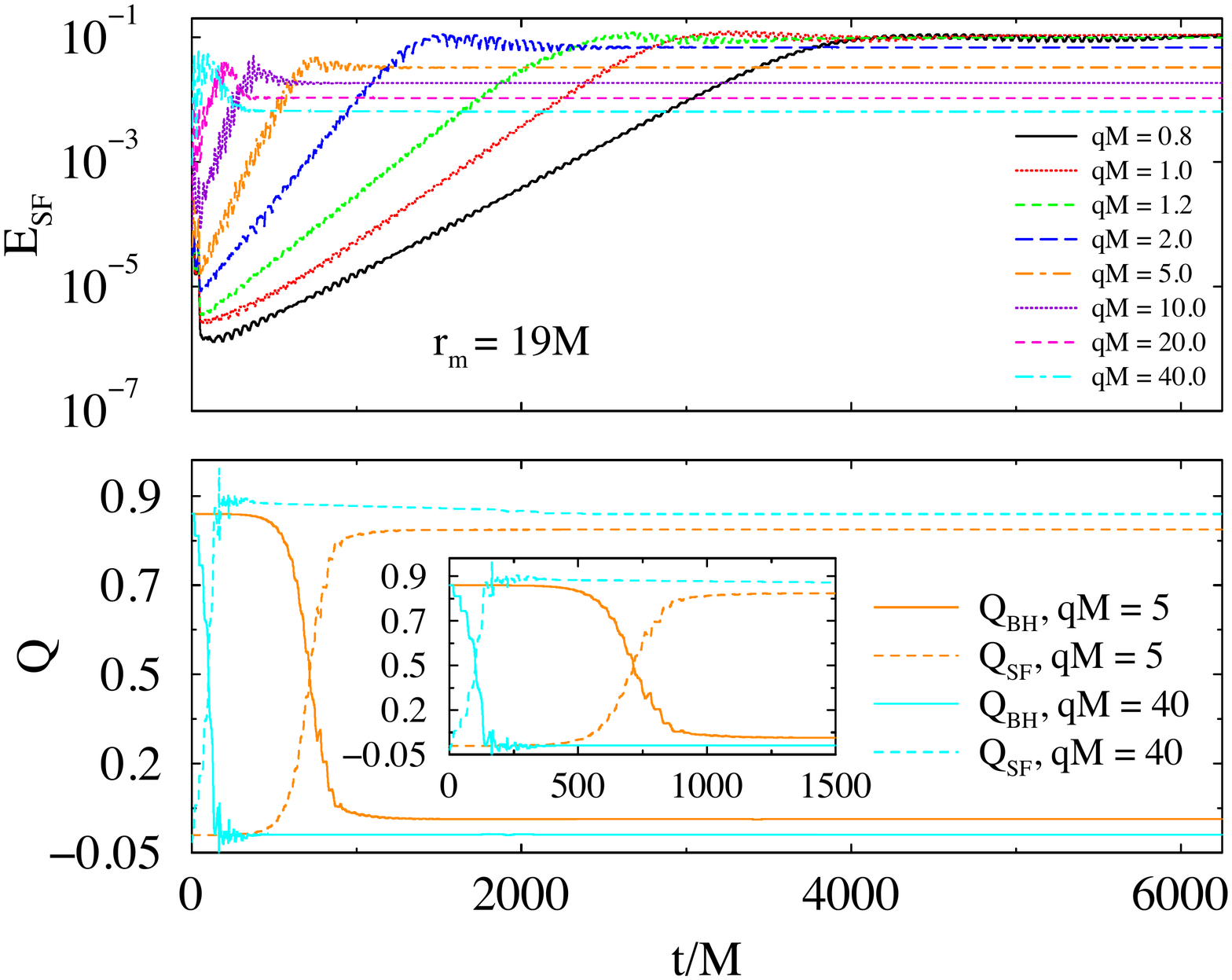} 
\caption{Time evolution of the scalar field energy and charge and the BH charge, displayed in logarithmic scale, for: (top panels) $r_{\rm{m}}=9M$; (middle panels) $r_{\rm{m}}=14.2M$; (bottom panels) $r_{\rm{m}}=19M$. The inset zooms in the early phase of the evolution, for clarity.}
\label{fg:esf_q_1}
\end{minipage}
\end{figure}

The first consequence of varying the mirror radius is a variation in the time scale of the process (for all other parameters fixed): the larger the mirror radius, the larger the e-folding time. This is an intuitive behaviour, as the recurrent scattering that leads to the exponential pile up of the superradiant modes takes longer in a larger cavity. 
This behaviour had already been noticed in linear studies~\cite{Herdeiro:2013pia}. 
Such trend is more easily visualized in Fig.~\ref{fg:esf_q_1}, where the time evolution of the scalar field is exhibited for the various values of $q$ and for the three values of the mirror radius.

Another clear trend when increasing the mirror radius is that the critical frequency at which equilibrium is achieved is smaller. Naively this is associated to a larger wavelength of the dominant superradiant mode, which is allowed in a larger cavity. A smaller critical frequency implies a smaller horizon electric potential and thus a larger charge to energy ratio transfer to the scalar field. This is in agreement with what can be observed from the table. Concerning the charge, the relevant information is in  the last two columns of Table I: for the same $q$, a larger radius implies a larger (smaller) fraction of charge in the scalar field (BH). Note that the corresponding panels of Fig.~\ref{fg:esf_q_1} show a perfect charge exchange, between the BH and the scalar field.  Concerning the energy transfer, inspection of the sixth to eighth column of Table I shows that, increasing the mirror radius, leads to a smaller energy growth of the scalar cloud. This inverse correlation between charge transfer and energy transfer had already been observed in~\cite{Sanchis-Gual:2015lje} and also occurs when varying $q$. Increasing the scalar field charge  (likewise increasing the mirror radius)  leads to a higher charge transfer to the scalar field but lower energy growth of the scalar field cloud. In terms of the strength of the instability, however, measured by the e-folding time, increasing the scalar field charge leads to the opposite trend to that of increasing the mirror radius: a larger scalar field charge leads to a faster growth of the instability.

\subsubsection{Impact of scalar field mass}
\label{sec_mass}

In our simulations presented in~\cite{Sanchis-Gual:2015lje} we chose to discuss a massive scalar field, as it seems far-fetched to consider a massless, but charged, scalar field (all charged particles are massive, in the Standard Model of particle physics). Still, for the sake of completeness, we here discuss the effect of the mass, by comparing simulations of a massive ($\mu M=0.1$) and a massless scalar field, and focusing on a particular feature of the field distribution in the equilibrium state.

\begin{figure}[h!]
\begin{center}
\includegraphics[width=0.4\textwidth]{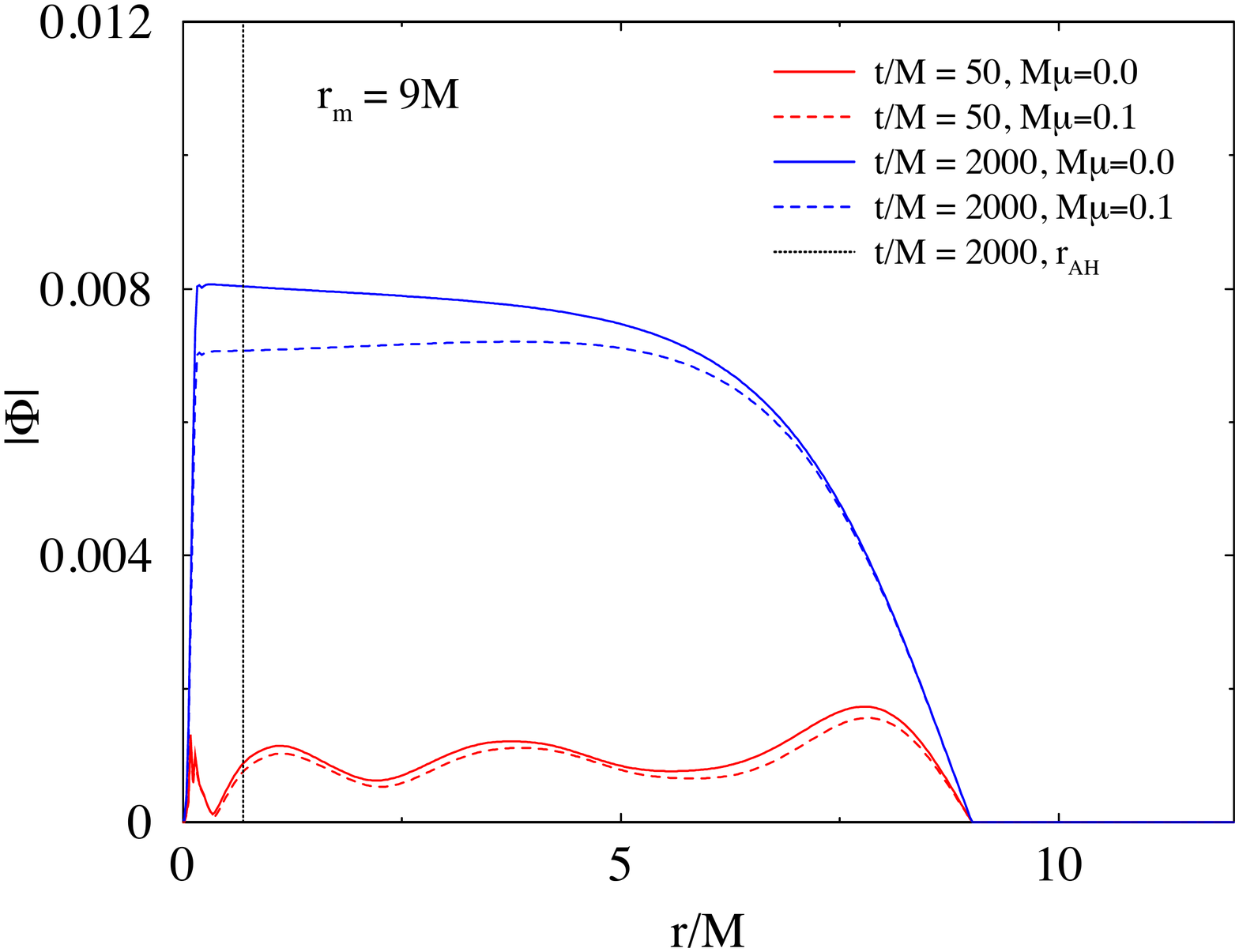}\vspace{-0.5cm}\\
\includegraphics[width=0.4\textwidth]{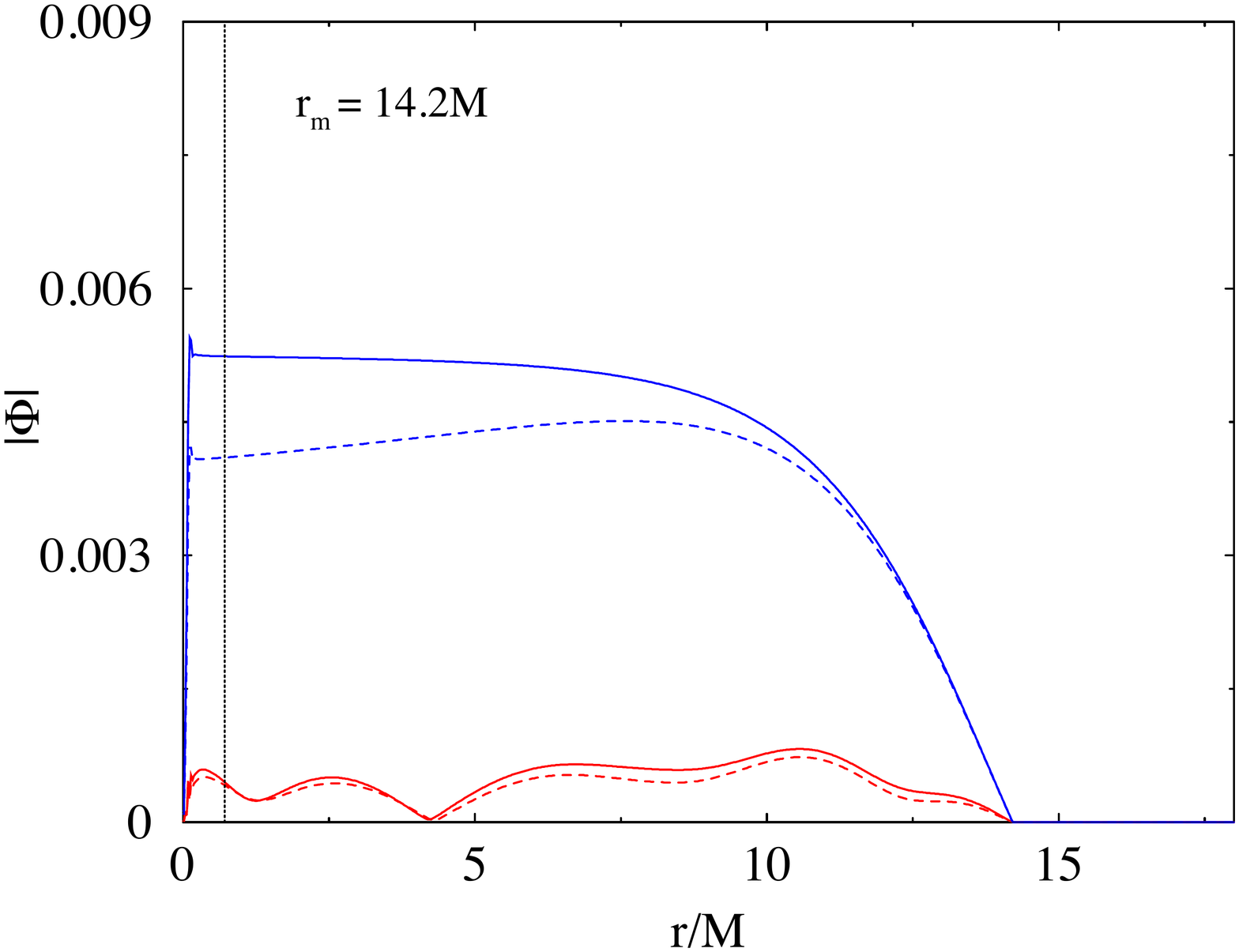}\vspace{-0.5cm}\\
\includegraphics[width=0.4\textwidth]{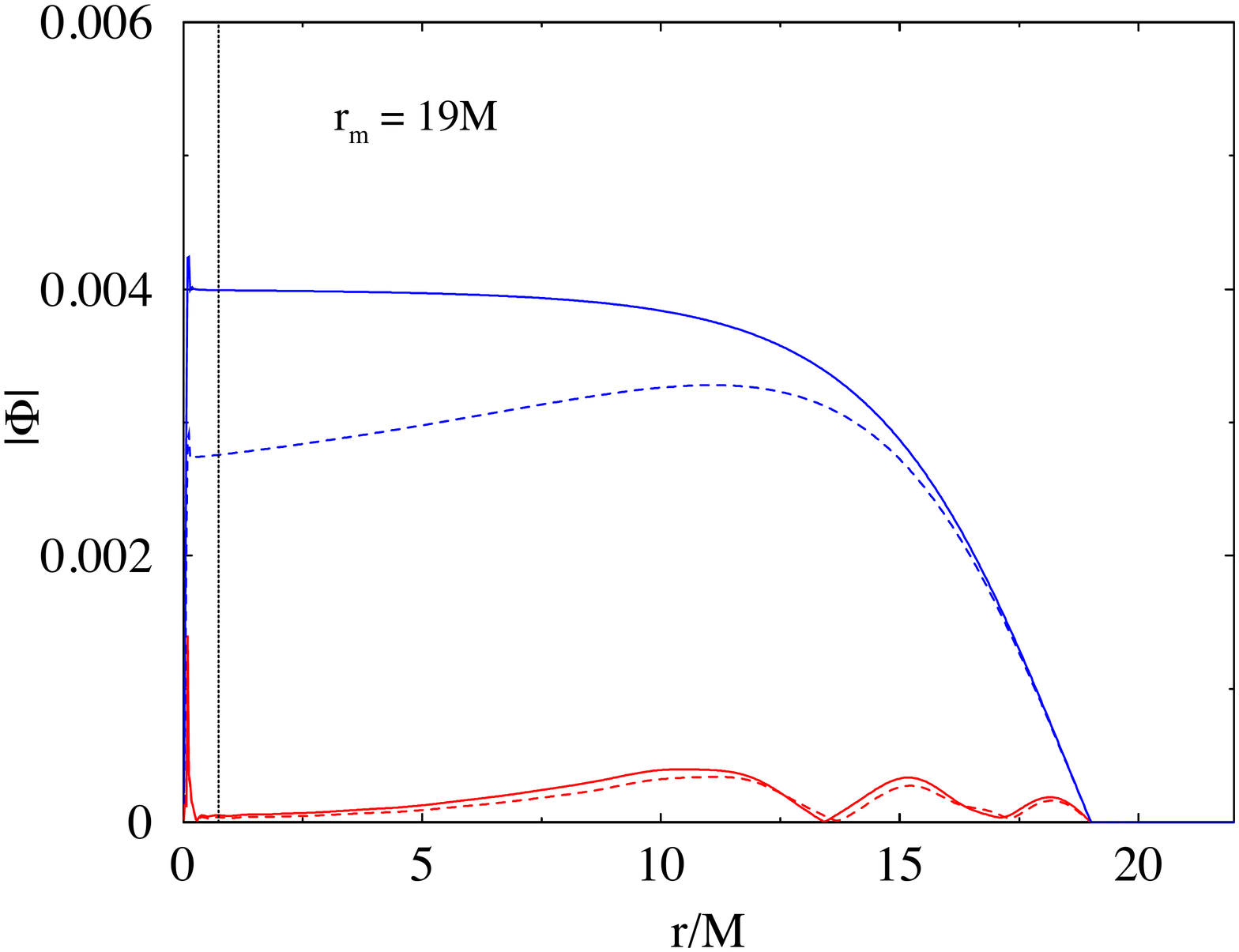}
\caption{Scalar field magnitude at two different time slices, for two different values of the scalar field mass, in terms of the radial coordinate, for $qM$ = 20 and $r_{m} = 9M$ (top panel), $r_{m}$ = $14.2M$ (central panel), $r_{m}=19M$ (bottom panel). The vertical line marks the location of the AH at the final time.}
\label{radialprofile}
\end{center}
\end{figure}

 In Fig.~\ref{radialprofile} we plot the scalar field magnitude, at two different time slices, for the evolution of the massive and the massless scalar field. 
 As can be observed from the various panels, at the first time slice plotted, $t=50M$, the scalar field distribution is ``bumpy", with several maxima and minima, and possibly with nodes. In the final time slice, however, $t=2000M$, corresponding to a late time at which equilibrium has been attained, there are no nodes. Moreover, whereas for the massless case the scalar field magnitude profile is monotonically decreasing from the horizon to the mirror, for the massive case there is a maximum.

Charged hairy BHs in a cavity at the threshold of the superradiant instability were constructed in~\cite{Dolan:2015dha}, for the model~\eqref{model} with $\mu=0$. Therein it was established that, amongst the different families of such hairy BHs, with different numbers of nodes for the scalar field magnitude between the horizon and the mirror,  only the nodeless solutions  are stable against perturbations (and hence could be the true end-point of the instability process). This is exactly what we find for our hairy BHs -- the scalar field magnitude is nodeless when equilibrium is reached. We remark that the stationary solutions in~\cite{Dolan:2015dha} were obtained for a massless scalar field; consequently the scalar field magnitude for the stable solutions was monotonically decreasing from the BH to the mirror, in agreement with what is found \textit{dynamically} in our simulations, and exhibited in Fig.~\ref{radialprofile}.

\subsubsection{Impact of the initial cloud parameters}
\label{sec_varycloud}

\begin{figure}
\begin{minipage}{1\linewidth}
\includegraphics[width=1.0\textwidth, height=0.3\textheight]{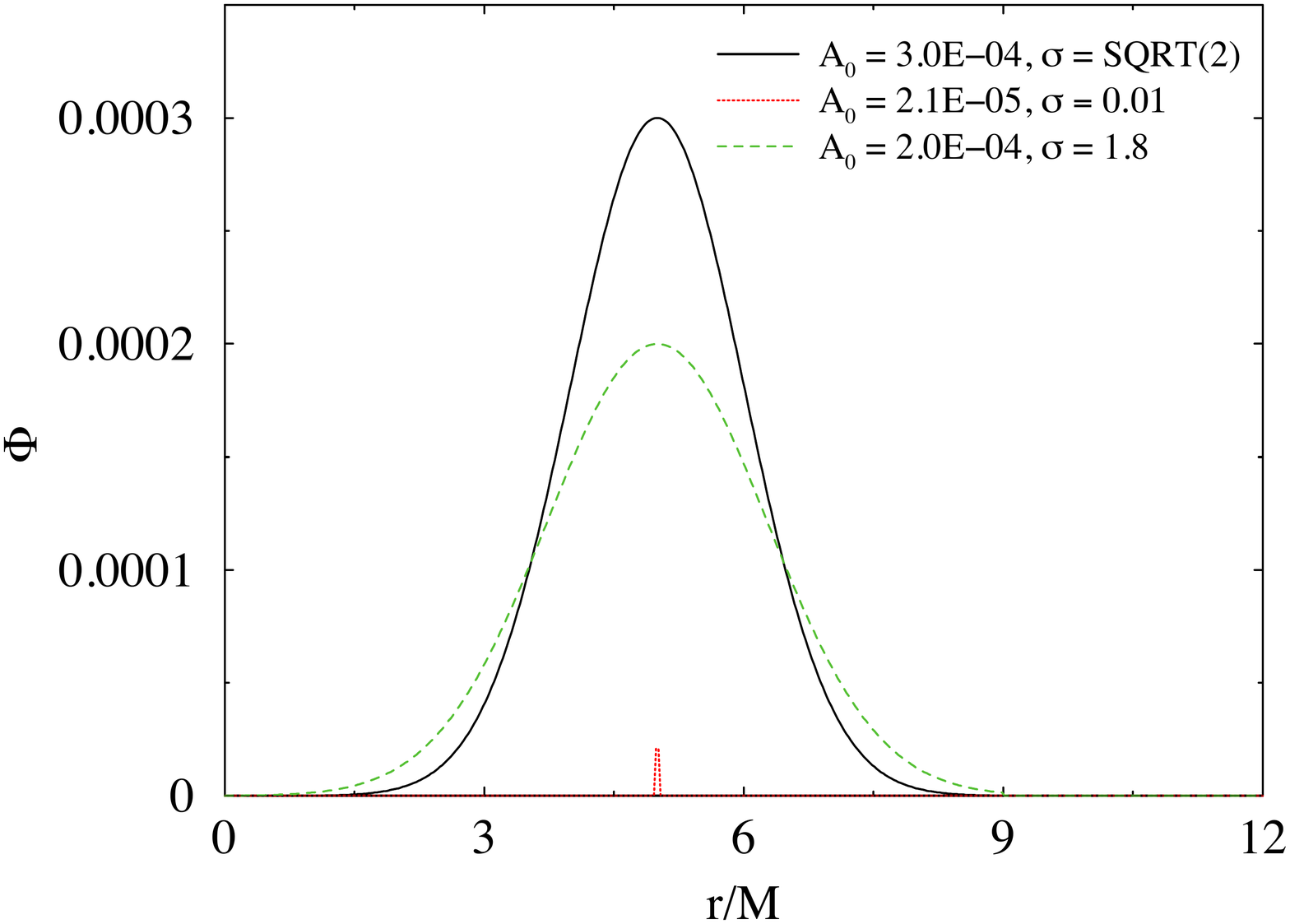} 
\includegraphics[width=1.0\textwidth, height=0.3\textheight]{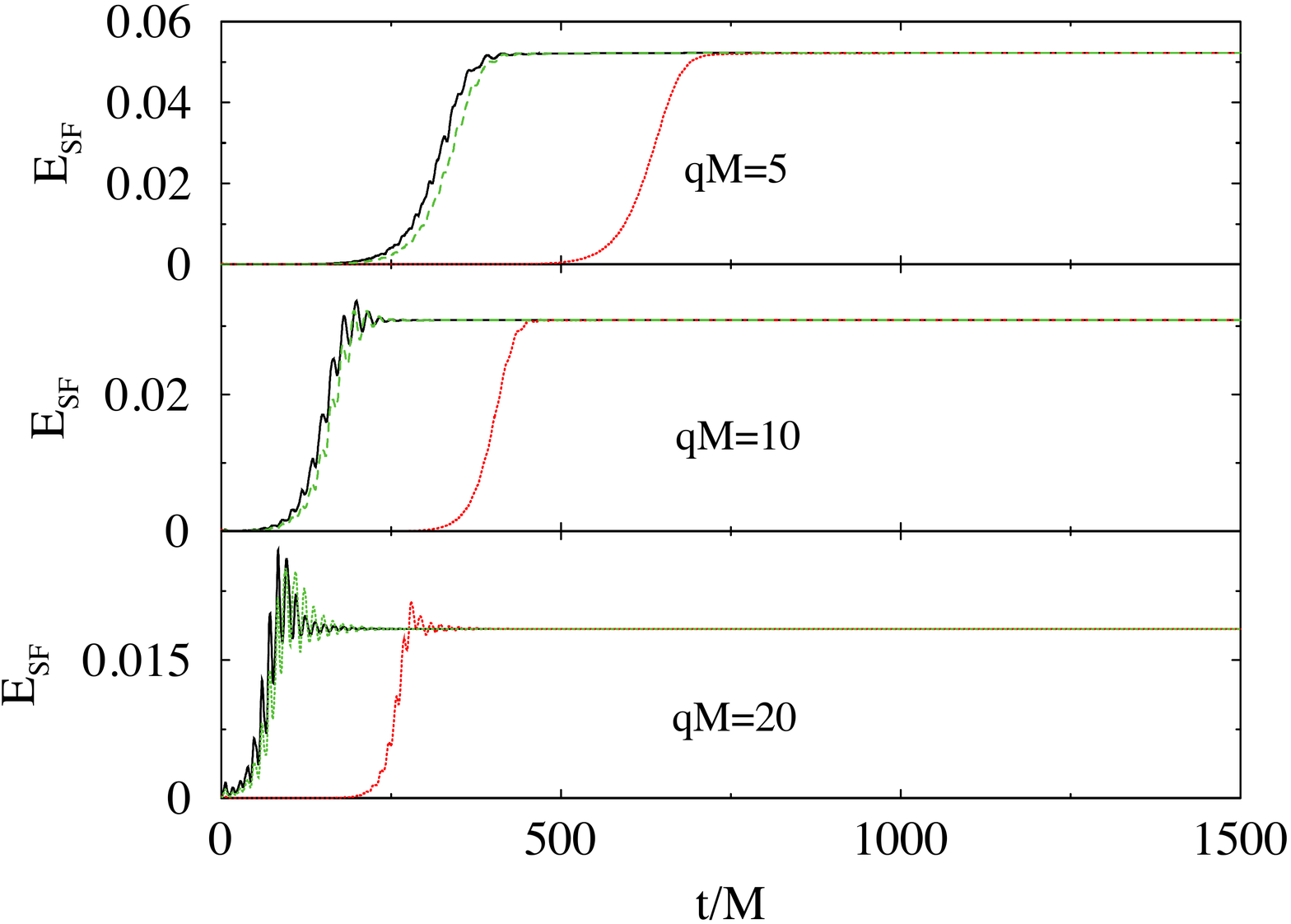} 
\caption{The three different Gaussians used as initial data (top panel). 
The corresponding time evolutions, for three different values of $qM$ (bottom panels).}
\label{fg:gaussian}
\end{minipage}
\end{figure}

In Fig.~\ref{fg:gaussian} we investigate the dependence of the evolution on the initial scalar perturbation. 
We compare three different perturbations. The black solid line corresponds to the default Gaussian, used in all 
other simulation presented in this paper ($A_0=3\times 10^{-4}$, $\sigma=\sqrt{2}$); the green dashed line 
corresponds to a scalar perturbation with a lower amplitude but slightly more spread $(A_0 = 2.0\times 10^{-4},\sigma=1.8)$; 
finally the red dotted line corresponds to a much lower amplitude perturbation and very narrow $(A_0 = 2.1\times10^{-5}, \sigma=0.01)$.  
The corresponding Gaussians are plotted, for comparison, in the top panel of Fig.~\ref{fg:gaussian}. 
The bottom panel shows the corresponding time evolutions of the scalar field energy, using the same color convention, 
from which one can extract three observations. Firstly, smaller 
perturbation amplitudes lead to a longer superradiant growth phase. Secondly, the final scalar field energy is 
insensitive to the initial perturbation. Thirdly, the scalar field energy overshoot (see Sec.~\ref{sec_bosenova} 
for a discussion of this overshooting behaviour) observed in the $qM=10,20$ cases is larger for larger perturbations. 
These features can be interpreted as the need to attain a certain threshold in the scalar cloud energy for the 
saturation phase to kick in. Naturally this threshold takes longer, when starting with a smaller perturbation. 
Still, the final hairy BH obtained is essentially insensitive to the perturbation parameters, as long as the perturbation 
approximation remains valid.

\begin{figure}[h!]
\begin{minipage}{1\linewidth}
\includegraphics[width=0.95\textwidth]{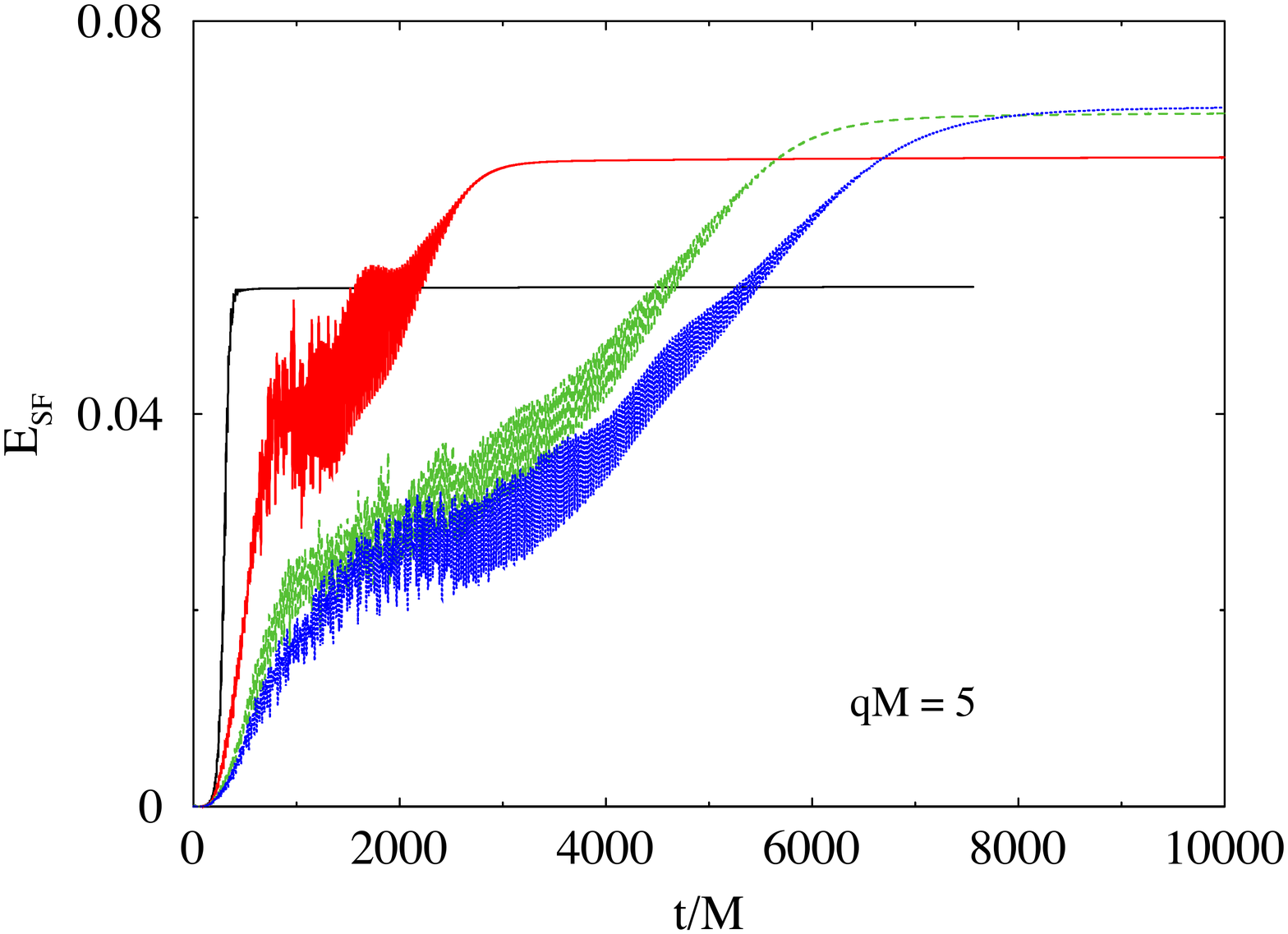}\vspace{-0.4cm}
\includegraphics[width=0.95\textwidth]{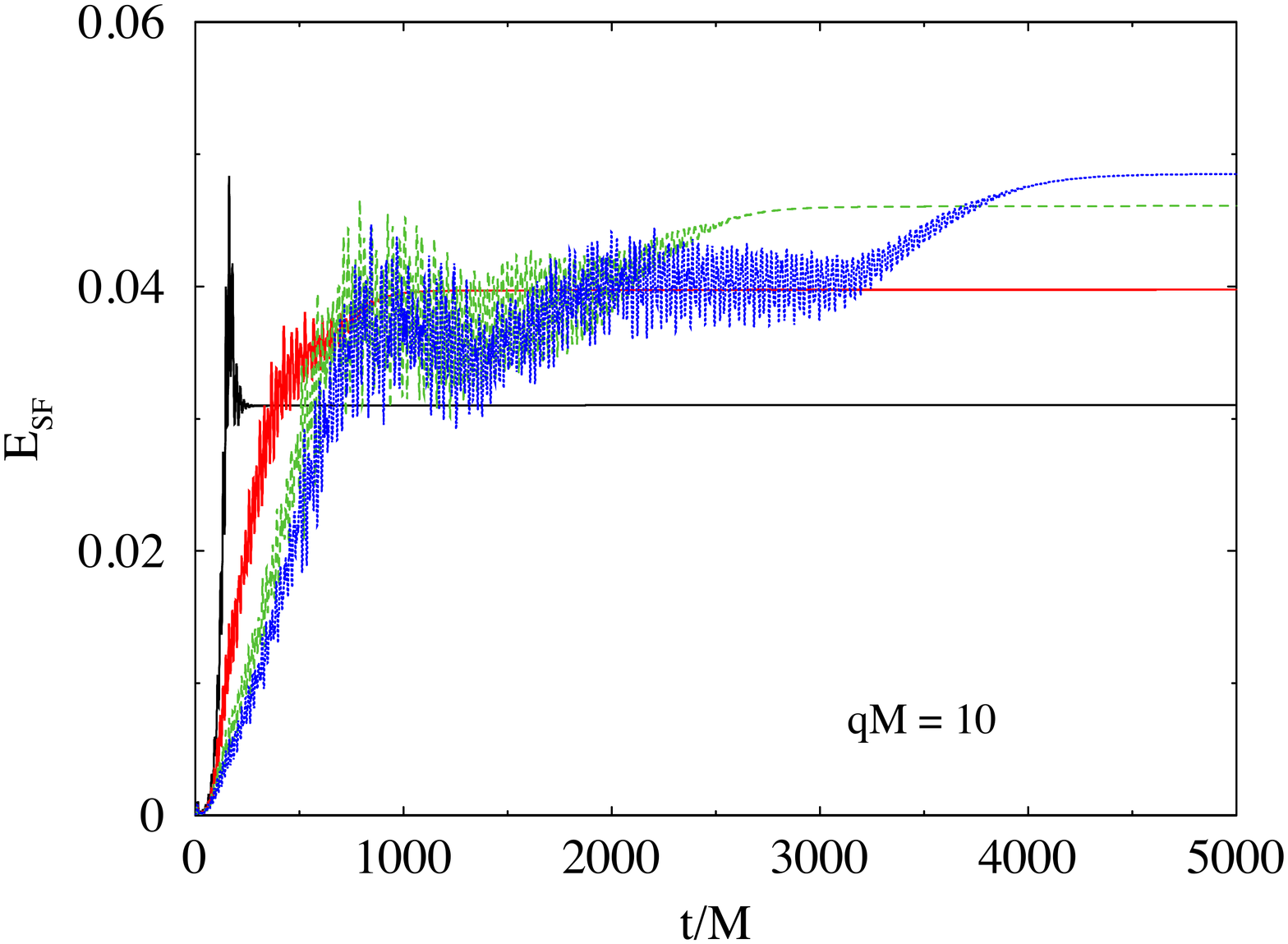}\vspace{-0.6cm}
\includegraphics[width=0.95\textwidth]{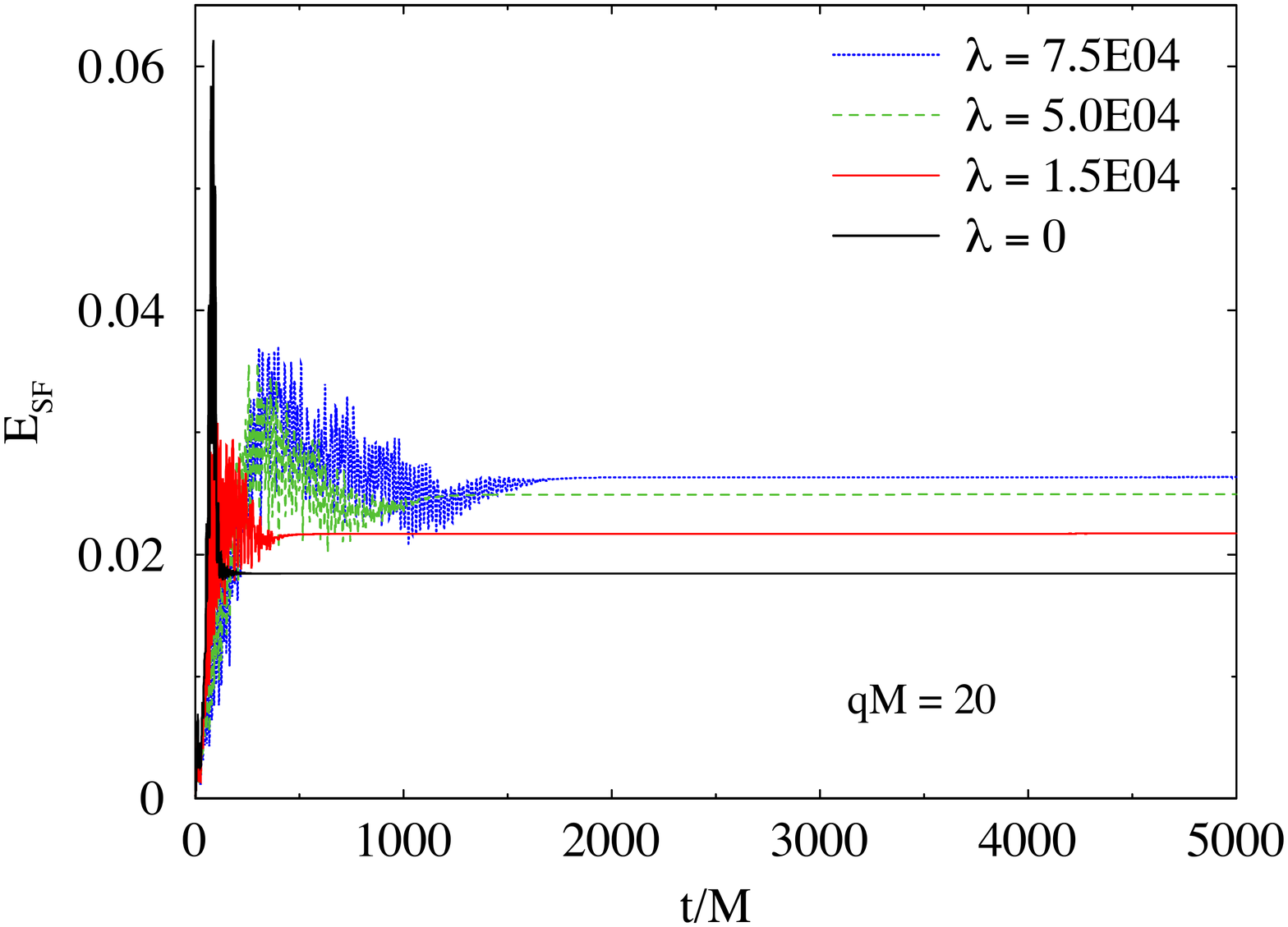}  
\caption{Time evolutions of the scalar field energy for different values of the quartic self-coupling and $qM=5,10,20$ (top, middle and bottom panels).}
\label{fg:nonlinear}
\end{minipage}
\end{figure}

\subsubsection{Impact of the scalar field self-interactions}
\label{sec_selfint}

We now tackle the effect of adding a quartic self-interaction to the scalar field,  
by taking $\lambda\neq 0$ in the model described by action~\eqref{model}. 
In Fig.~\ref{fg:nonlinear} we show the time evolution of the scalar field energy for three non-zero values of 
the quartic self-coupling together with the case with no self-interactions, for three different values of $qM$.

The overall trends revealed by inspection of Fig.~\ref{fg:nonlinear} is as follows. 
Increasing the self-coupling leads to a slower growth of the scalar field energy outside 
the horizon. But the final state corresponds to a hairy BH with more energy in the scalar 
field. Moreoever, the self-interactions promote more energy exchange between the BH and the 
scalar field outside the horizon, $i.e.$, the evolution is never monotonic, even for small $qM$ values. 
This is likely associated with the mode conversion allowed by the self-interactions, a suggestion 
supported by the mode analysis discussed below, in Sec.~\ref{sec_modes_nl}.

\begin{figure}[h!]
\begin{minipage}{1\linewidth}
\includegraphics[width=1.0\textwidth]{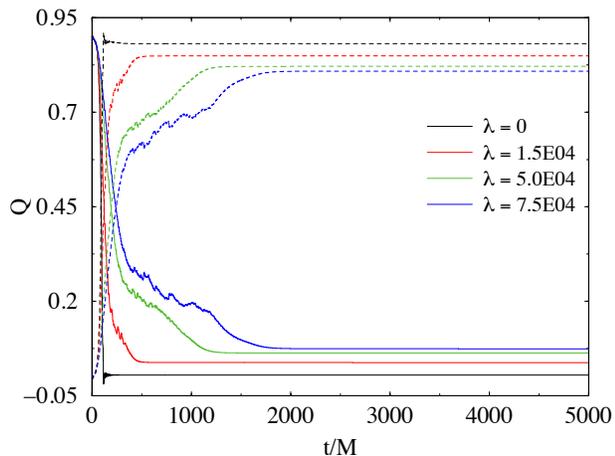} 
 \caption{Time evolution of the electric charge in the scalar field outside the horizon, 
 for different values of the quartic self-coupling and $qM=20$.}
\label{fg:nonlinearQ}
\end{minipage}
\end{figure}

As in all previous cases, the increase in the ability to transfer energy from the BH into the scalar 
field is accompanied by a decrease in the ability to transfer charge from the BH to the scalar field. 
This is illustrated in Fig.~\ref{fg:nonlinearQ} for the simulations with $qM=20$.

\begin{figure}[h!]
\begin{minipage}{1\linewidth}
\includegraphics[width=1.0\textwidth]{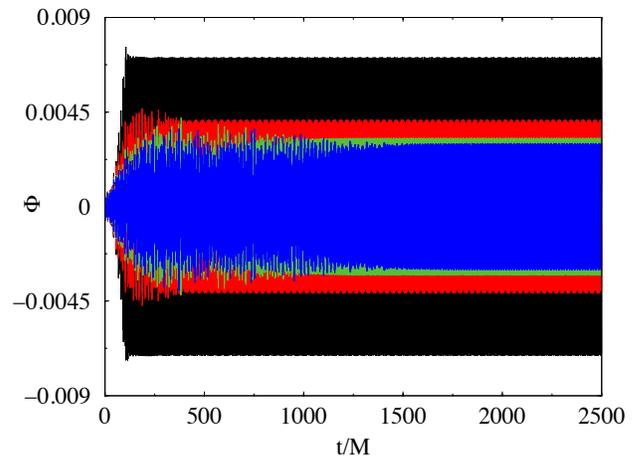} 
 \caption{Time series for the (real part of the) scalar field for different values of the quartic self-coupling (same color coding as in Fig.~\ref{fg:nonlinearQ}) and $qM=20$.}
\label{fg:nonlinearobs}
\end{minipage}
\end{figure}

\begin{figure}[h!]
\begin{minipage}{1\linewidth}
\includegraphics[width=1.0\textwidth]{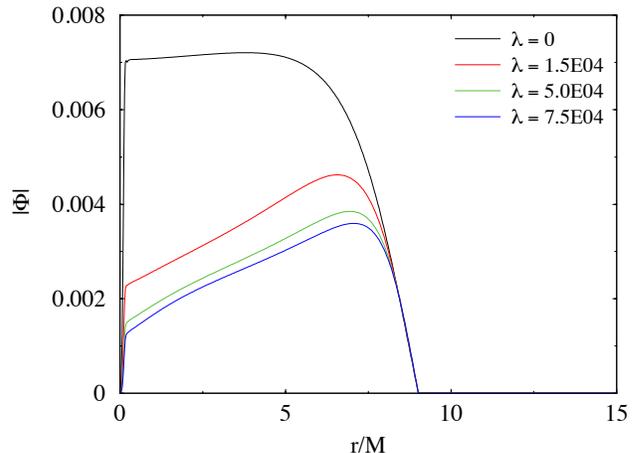} 
 \caption{Magnitude of the scalar field, in terms of the radial coordinate, for the final 
 BH configuration, for different values of the quartic self-coupling and $qM=20$.}
\label{fg:nonlinearprofile}
\end{minipage}
\end{figure}

Interestingly, the larger scalar field energy obtained for larger self-couplings is not 
associated with a larger scalar field amplitude outside the horizon. This can be concluded 
from Fig.~\ref{fg:nonlinearobs}, where the oscillations of the (real part of the) scalar 
field are shown for $qM=20$. It can be observed these oscillations are larger for 
\textit{smaller} self-coupling. This result is confirmed in Fig.~\ref{fg:nonlinearprofile}, 
where the magnitude of the final scalar field profile is shown as a function of the radial 
coordinate. This figure clarifies, moreover, that the scalar field \textit{spatial gradients} 
become larger when increasing the self-coupling. Thus, the larger gradients, rather than a 
larger scalar field magnitude, yield the larger scalar field energy outside the horizon, for larger self-coupling.

As we saw before, the presence of a mass term leads to an extremum in the scalar field 
magnitude radial profile (as opposed to a monotonic function for the massless case, $cf.$ 
Sec.~\ref{sec_mass}), and hence a larger radial second derivative of that magnitude. 
The self-interactions term, from Fig.~\ref{fg:nonlinearprofile}, tends to further increase 
this second derivative, in the neighbourhood of the extremum.

\subsection{Bosenova and mode analysis}
\label{sec_bosenova}
\begin{figure}
\begin{minipage}{1\linewidth}
\includegraphics[width=1.0\textwidth, height=0.3\textheight]{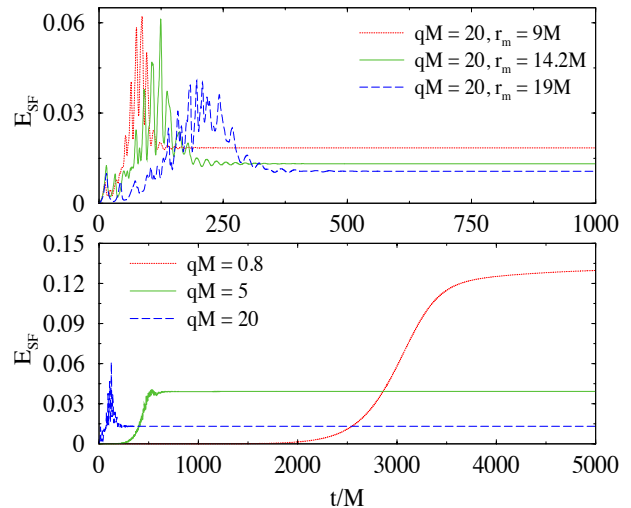} 
\caption{Top panel: Details of the oscillations of the energy density during the ``explosive'' phase, for  
the $qM =20$ models and three different positions of the mirror. The extracted energy overshoots 
the final equilibrium value, and strong oscillations follow. Bottom panel: Variation of the scalar 
field energy density for three models for which the mode analysis is performed in Fig.~\ref{fg:modes}.}
\label{fg:bosenova}
\end{minipage}
\end{figure}

Analysis of Fig.~\ref{fg:esf_q_1} reveals a qualitative difference in the evolution of the scalar field energy between low and high scalar field charge simulations. Whereas the former exhibit an essentially monotonic growth, the latter display a more turbulent evolution before the equilibrium phase, wherein the energy extraction overshoots the equilibrium value and some energy is returned to the BH. This behaviour is detailed in Fig.~\ref{fg:bosenova} (top panel) for  $qM=20$ and for the three different positions of the mirror. The figure shows strong oscillations in the scalar field energy contained outside the horizon, before the system relaxes into an equilibrium configuration. Observe also that when the mirror is set closer to the BH, the relaxation is faster. 

\begin{figure}[h!]
\begin{minipage}{1\linewidth}
\includegraphics[width=0.9\textwidth, height=0.2\textheight]{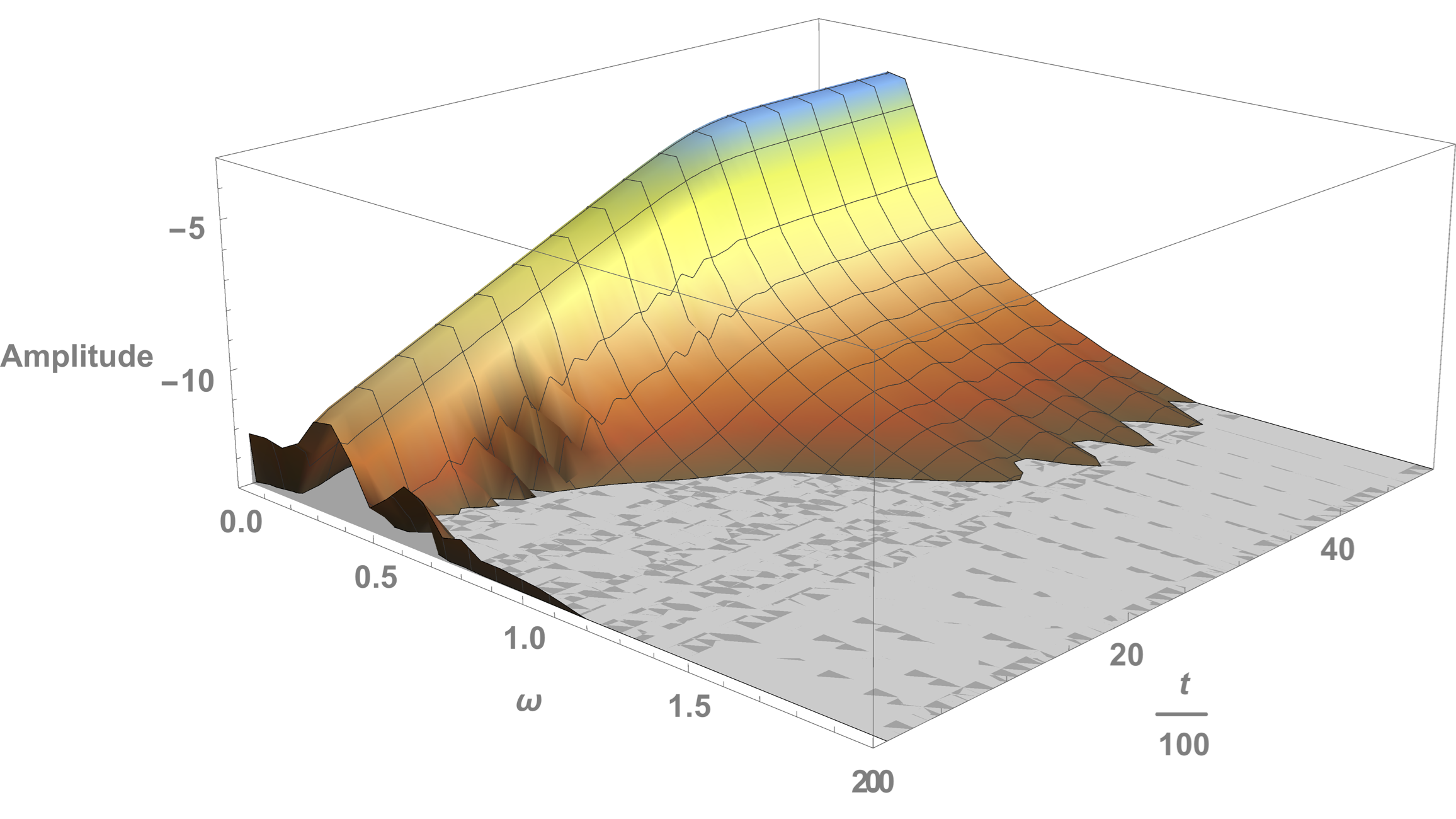}\\
\includegraphics[width=0.9\textwidth, height=0.2\textheight]{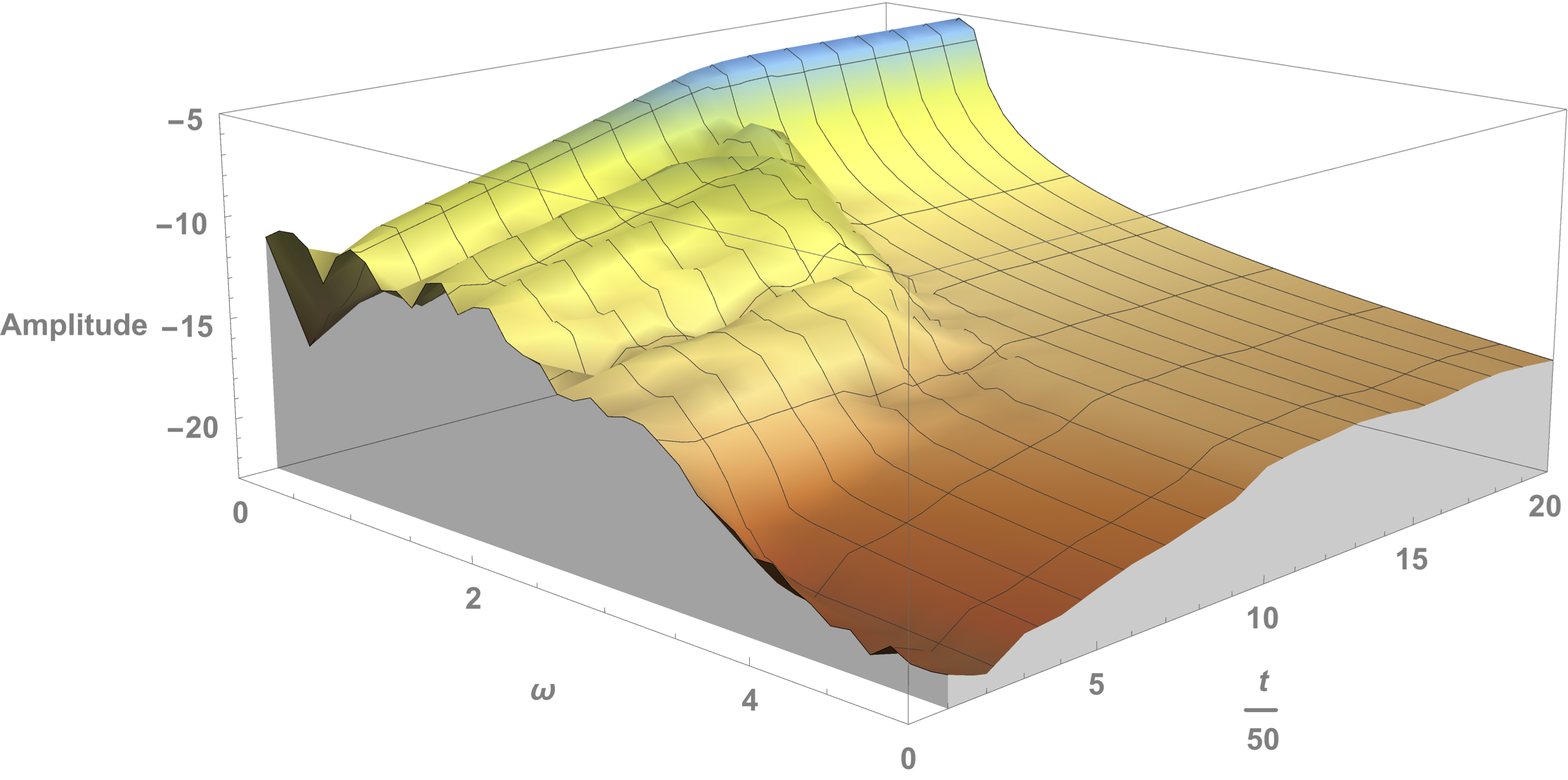}\\
\includegraphics[width=0.9\textwidth, height=0.2\textheight]{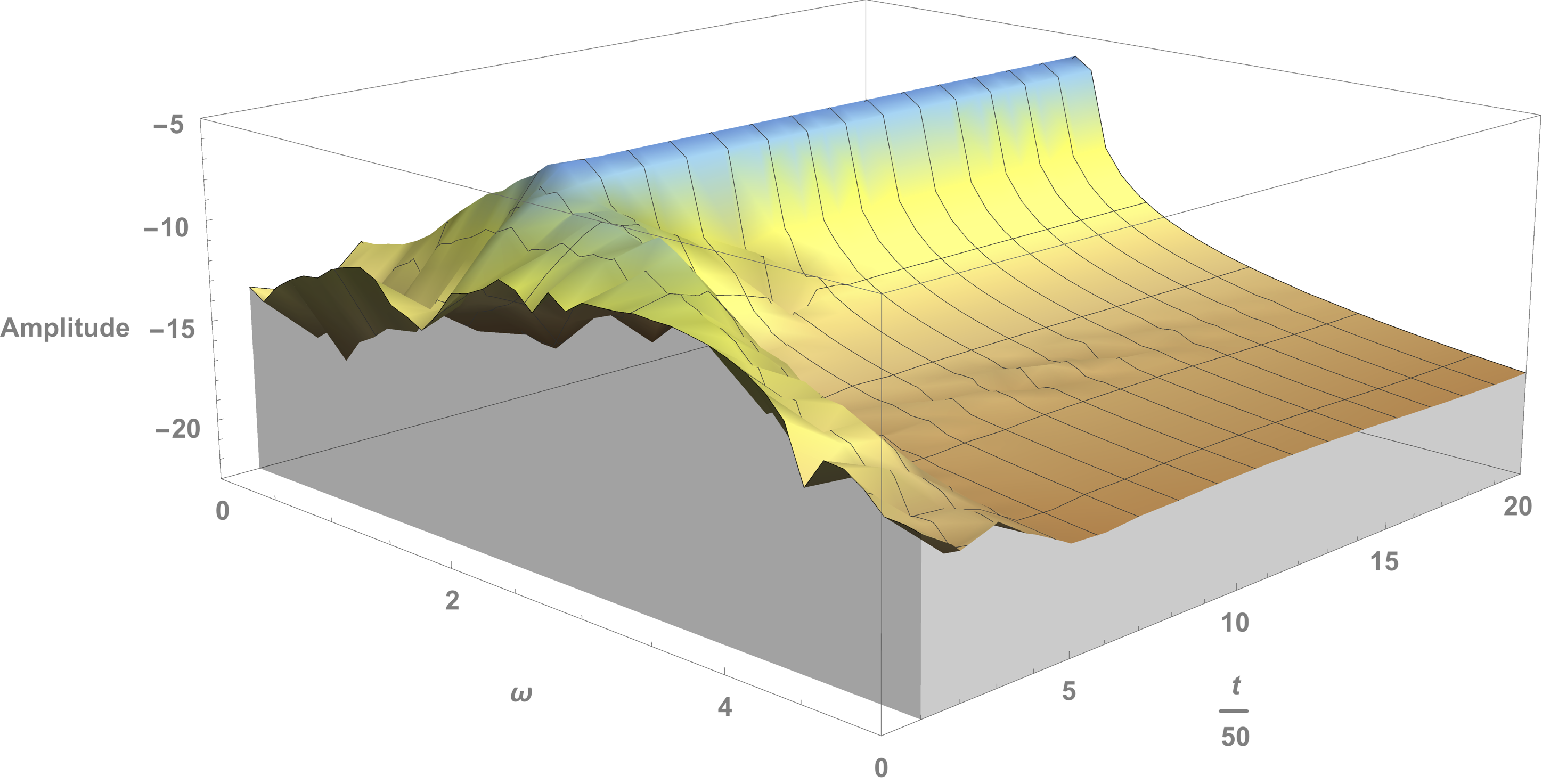} 
\caption{Mode analysis for (top panel): $qM=0.8$, (central panel) $qM=5$, (bottom panel) $qM=20$. For all three cases $\mu M=0.1$ and $r_m=14.2M$.}
\label{fg:modes}
\end{minipage}
\end{figure}

During the oscillations observed in Fig.~\ref{fg:bosenova}, some of the energy in the scalar field is pushed back into the BH,  before being extracted again, in a process that can last several cycles.  In~\cite{Sanchis-Gual:2015lje}, it was suggested this process resembles the bosenova explosion, described in \cite{Yoshino:2012kn, Yoshino:2015nsa}. Such explosion, resulting from the non-linear interactions of the scalar field, would push the energy  of a test, but non-linear, scalar field on the Kerr background, back into the BH. A simpler explanation, moreover not needing to invoke non-linear effects, was put forward in~\cite{Bosch:2016vcp}, by studying the growth of the superradiant instability in charged AdS BHs, a setup with analogous physics to the one studied herein. These authors argued that oscillations such as the ones observed in Fig.~\ref{fg:bosenova} result from modes that become non-superradiant, as the horizon electric potential (and hence the critical frequency) decreases, and consequently fall back into the BH. In order to test this hypothesis in our setup, we have performed the mode analysis shown in Fig.~\ref{fg:modes}.  This figures shows that for $q$ just above the instability threshold (the smallest $q$ value, $qM=0.8$), the system only has a single superradiant mode, and the evolution consists of a very smooth transition to the stationary equilibrium state, in agreement with the low $q$ curves in Fig.~\ref{fg:esf_q_1}. For larger $q$ ($qM=5$ and 20), one observes more than one initially superradiant mode growing, since they are in the superradiant range, but decay before the end state is reached, as they exit the superradiant window. This qualitatively explains the oscillations seen in Fig.~\ref{fg:bosenova}.  In this case, the only mode that does not decay is the fundamental mode, which matches the critical frequency as the system relaxes into the hairy BH solution.

\subsubsection{Mode analysis with $\lambda\neq 0$}
\label{sec_modes_nl}

\begin{figure}
\begin{minipage}{1\linewidth}
\includegraphics[width=1.0\textwidth]{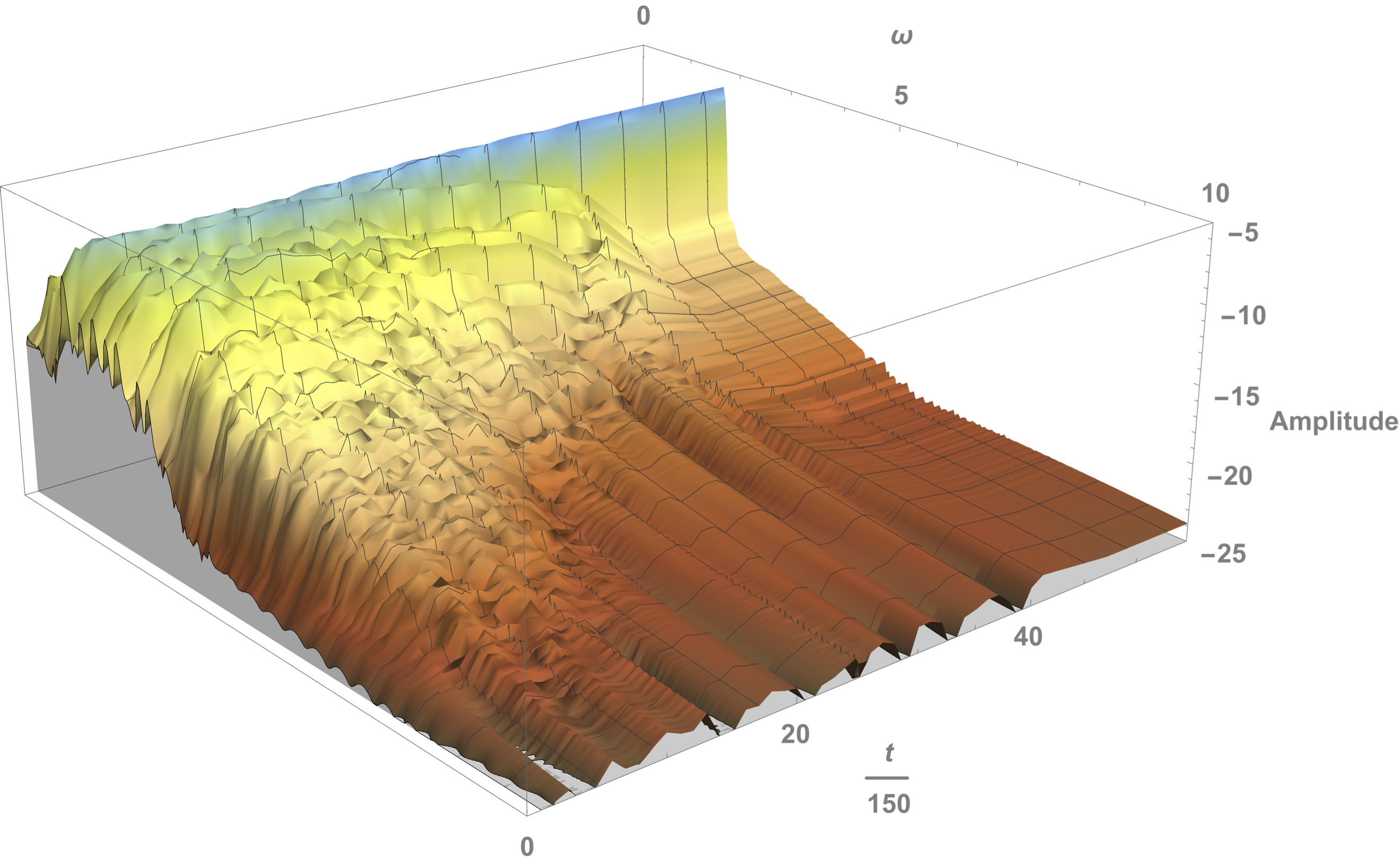} 
\includegraphics[width=1.0\textwidth]{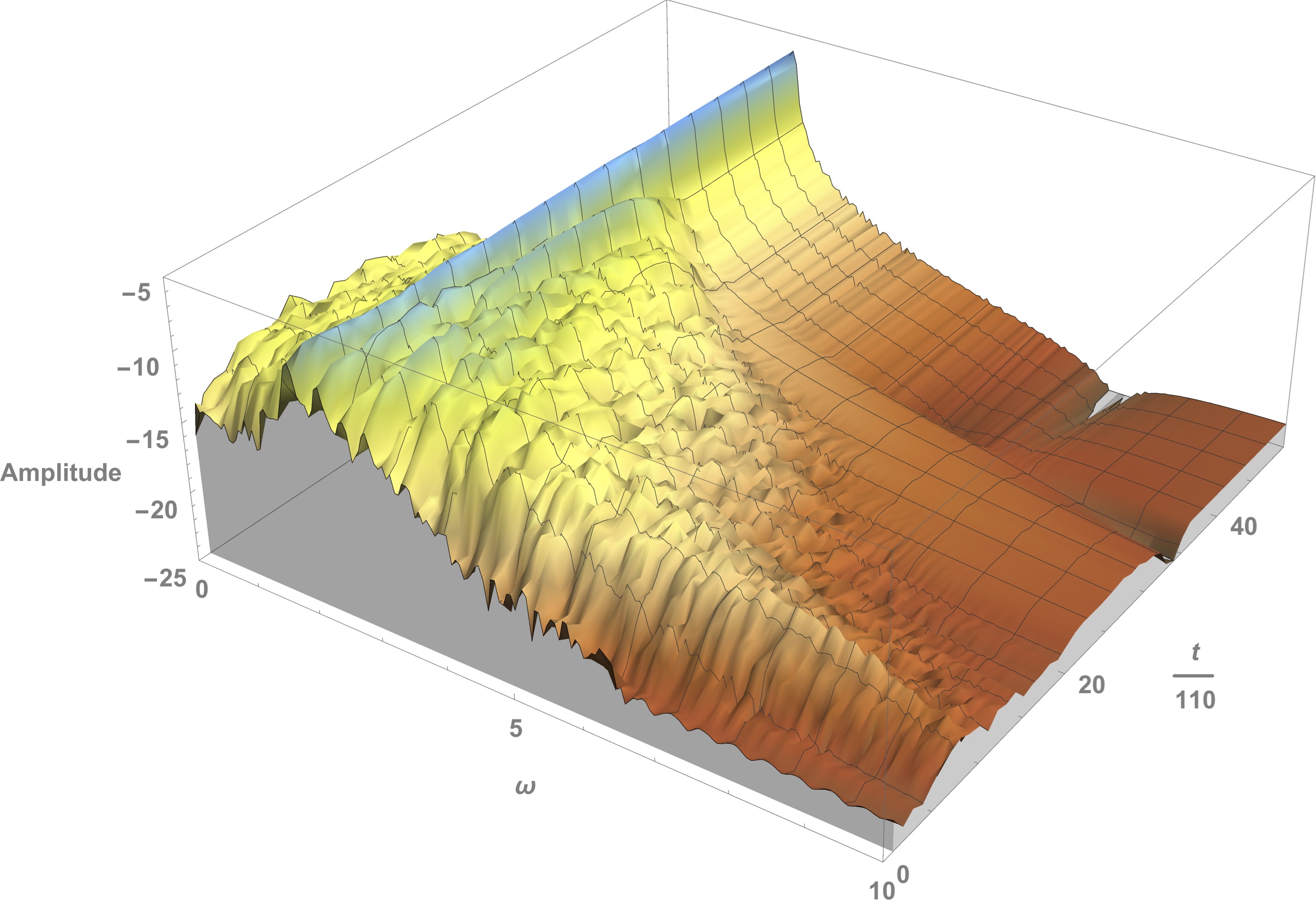} 
\includegraphics[width=1.0\textwidth]{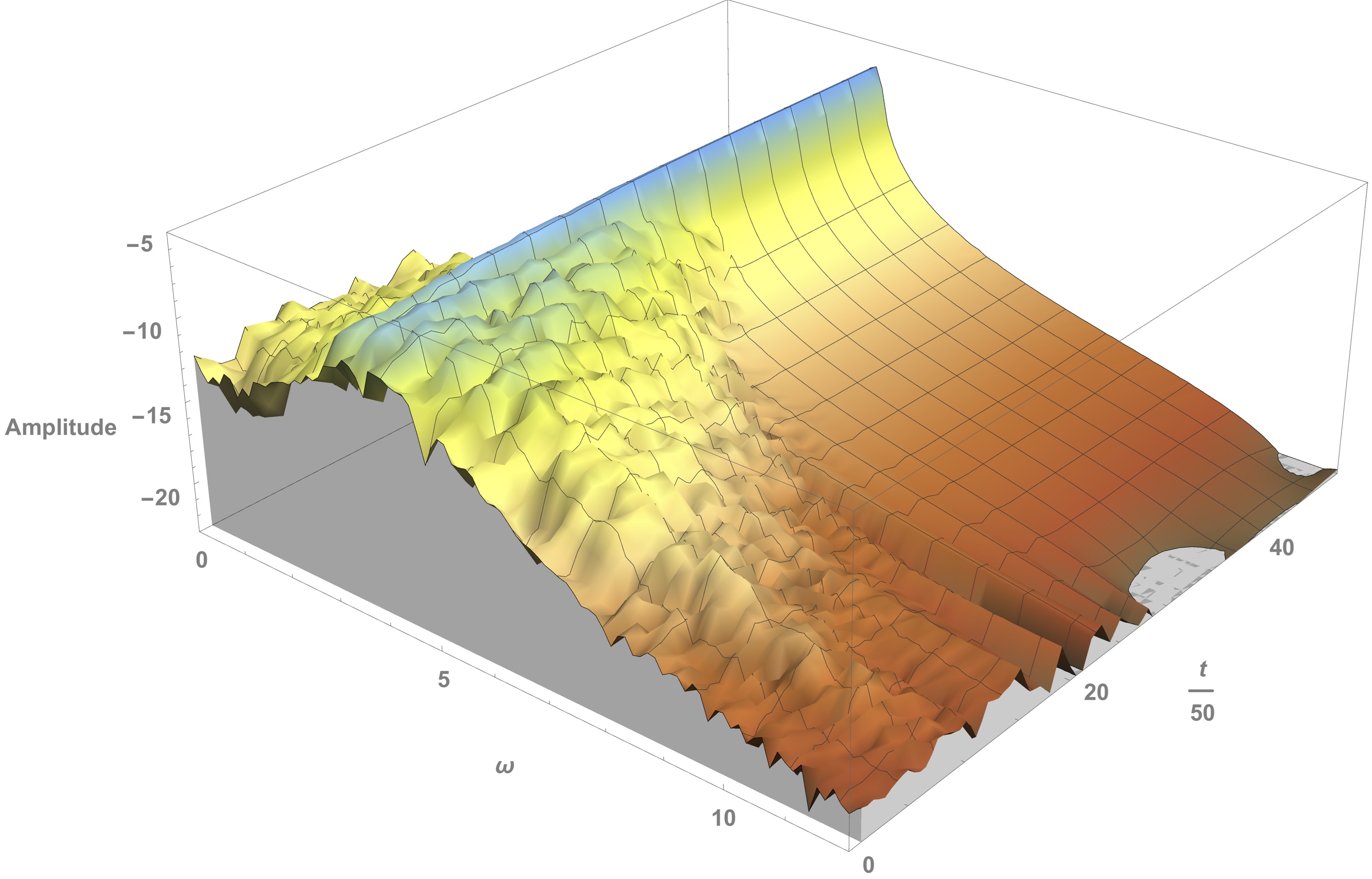} 
\caption{Mode analysis for the model with self-interactions, with $\lambda=7.5\times 10^4$ and: 
(top panel) $qM=5$, (central panel) $qM=10$, (bottom panel) $qM=20$. For all three cases $\mu M=0.1$ and $r_m=9M$.}
\label{fg:modesnl}
\end{minipage}
\end{figure}

The mode analysis of the previous subsection suggests that despite the non-linear nature of 
the process leading to the hairy BH formation, different scalar field modes evolve in an 
essentially independent way and, moreover, in the way predicted by the linear (test field) 
theory. A natural question is how the scalar field self-interactions affect such evolution. 
To address this question we plot in Fig.~\ref{fg:modesnl} a mode analysis for the evolution 
with $\lambda=7.5\times 10^4$ and $qM=5,10,20$. Some differences with respect to the cases 
without self-interactions shown in Fig.~\ref{fg:modes} are notorious. A first difference, 
is that the dominant mode, that ends up defining the final BH hair, is essentially unchanged 
during the evolution. In particular the growth phase, expected from linear theory, is suppressed. The reason is that for the (large) values of $\lambda$ (and small mass $\mu$) considered, the self-interacting (quartic) term is almost of the order of the (quadratic) mass term from the very beginning and hence the linear approximation never holds. A second difference, is that the remaining modes, that end up decaying into the BH, are now more 
turbulent. It is plausible that this is a manifestation of mode conversion, promoted by the 
self-interactions. Of course, such mode conversion can also occur, even without the manifest 
scalar self-coupling, due to the implicit self-coupling induced via the coupling to gravity. 
Nonetheless, our findings are that, for the setups and parameters considered herein, the effect 
is clearer in the presence of a non-vanishing self-interaction term.

\section{Conclusions} 
\label{sec:conclusions}

In this paper we have extended and complemented the results presented in a recent letter~\cite{Sanchis-Gual:2015lje} on the non-linear development of the superradiant instability for a RN BH in a cavity. Following the development of this instability, we have shown it leads to the dynamical formation of a hairy RN BH, of the type studied in~\cite{Dolan:2015dha} as stationary solutions. This falsifies the weak version of the no-hair hypothesis, albeit not for a truly asymptotically flat spacetime. In Fig.~\ref{fig_evolution} we provide an illustration of the dynamical formation of the hairy BH.

We have examined the sensitivity of the hair growth process to the BH charge, the mirror radius, the scalar charge and mass, 
the parameters of the initial scalar perturbation and to the introduction of a scalar self-coupling. In a nutshell, 
the energy extraction is \textit{more} efficient for lower scalar field charge, for larger BH charge and for smaller mirror radius.  
The trend with the charge extraction is opposite:  it is \textit{less} efficient for lower charge coupling, for larger BH charge and 
for smaller mirror radius. Concerning the existence, or not, of  scalar field mass we have confirmed that this leads to a 
qualitative difference in the final scalar field magnitude profile, which is monotonically decreasing, from the horizon to 
the mirror, for massless scalar fields, but has a maximum for massive scalar fields. Introducing a scalar field quartic 
self-coupling, the final scalar field magnitude profile acquires larger spatial gradients, which justifies the larger energy 
transferred from the BH to the scalar field, despite the lower amplitude of the final scalar field profile, as compared to 
the non-self-interacting case. We have also observed that the final hairy BH is essentially insensitive to varying the 
initial perturbation, even though the details of the evolution depend on it.

We have clarified the oscillating behaviour for the scalar field energy outside the horizon which is observed for the 
larger scalar field charges. A mode analysis reveals that various modes contribute to the superradiant growth in the 
early states of the process. However, a single mode remains at the end, in equilibrium with the BH; thus the other modes became non-superradiant and decay back into the BH before equilibrium is attained. This is in contrast with the smaller $q$ simulations, for which a single mode is superradiantly growing from the early stages of the process and hence the equilibrium phase is achieved essentially monotonically. This analysis confirms the observations in~\cite{Bosch:2016vcp}, 
for our setup. Such mode analysis lends support to the linear approximation and even to the use of an adiabatic approximation,
such as in~\cite{Brito:2014wla}, for taking into account the backreaction. Indeed, individual modes evolve essentially independently, 
exchanging their energy with the horizon. When turning on self-interactions, however, the picture changes. For sufficiently large self-coupling, the regime 
predicted by the linear theory is essentially unobserved, and each mode, except the dominant one, fluctuates noticeably 
until it completely decays. Not surprisingly, therefore, turning on self-interactions limits the validity of a linear approximation.

Finally, we remark that the hairy BHs we have dynamically shown to form in this paper, can be interpreted as a bound state of a RN BH and a charged scalar soliton in a cavity. This latter class of solutions was recently studied in detail in~\cite{Ponglertsakul:2016wae}. It was shown in this work that, amongst these solitonic solutions, some are unstable. An interesting question is, thus, what is the development of the instability for such unstable solitons, and in particular, if they evolve into a hairy BH. The technology described herein can be used to tackle this question. We hope to report on it in the near future. 

\section*{Acknowledgements}

This work has been supported by the Spanish MINECO (AYA2013-40979-P), 
by the Generalitat Valenciana (PROMETEOII-2014-069), by the 
CONACyT-M\'exico grant No. 233137, by the Max-Planck-Institut f{\"u}r Astrophysik, by the FCT (Portugal) IF programme, by the CIDMA (FCT) strategic project UID/MAT/04106/2013 and by  the  H2020-MSCA-RISE-2015 Grant No.  StronGrHEP-690904. Computations have been 
performed at the Servei d'Inform\`atica de la Universitat de Val\`encia.

\appendix
\section{Source terms}
\label{appendix}

In this Appendix the source terms included in the explicit or partially implicit operators are detailed.

Firstly, $a$, $b$, $X=\psi^{-1/2}$, $\alpha$, $\beta^r$, $\Phi$ and $E^{r}$, are evolved explicitly, i.e., all the source terms of the evolution equations of these variables are included in the $L_1$ operator of the second-order PIRK method. 

Secondly, $A_a$ and $K$, are evolved partially implicitly, using updated values 
of $\alpha$, $a$ and $b$. More precisely, the corresponding $L_2$ and $L_3$ 
operators associated with the evolution equations for $A_a$ and $K$ read:
\begin{align}
	L_{2(A_a)} &= - \left(\nabla^{r}\nabla_{r}\alpha 
- \frac{1}{3}\nabla^{2}\alpha\right) 
+ \alpha\left(R^{r}_{r} - \frac{1}{3}R\right) \ , \\
	L_{3(A_a)} &= \beta^{r}\partial_{r}A_{a} + \alpha K A_{a} 
- 16\pi\alpha(S_a - S_b) \ , \\
	L_{2(K)} &= - \nabla^{2}\alpha \ , \\
	L_{3(K)} &= \beta^{r} \partial_{r}K  
+ \alpha(A_{a}^{2} + 2A_{b}^{2} + \frac{1}{3}K^{2}) \nonumber \\
& + 4\pi\alpha(\rho + S_{a} + 2S_{b}) \ .
\end{align}
Next, $\hat{\Delta}^{r}$, $\Psi$, $\Pi$, $\varphi$ and $a_{r}$ are evolved partially implicitly, using the updated values 
of $\alpha$, $a$, $b$, $\beta^r$, $\psi$, $A_a$, $K$, $\Phi$ and $E^{r}$.
Specifically, the corresponding $L_2$ and $L_3$ operators associated with the 
evolution equation for $\hat{\Delta}^{r}$, $\Psi$, $\Pi$, $\varphi$ and $a_{r}$ are given by:
\begin{align}
	L_{2(\hat{\Delta}^{r})} &= \frac{1}{a}\partial^{2}_{r}\beta^{r} 
+ \frac{2}{b}\partial_{r}\left(\frac{\beta^r}{r}\right)
+ \frac{\sigma}{3 a}\partial_{r}(\hat{\nabla}_m\beta^{m}) \nonumber \\
  & - \frac{2}{a}(A_{a}\partial_{r}\alpha + \alpha\partial_{r}A_{a}) 
- \frac{4\alpha}{r b}(A_{a}-A_{b}) \nonumber \\
	& + \frac{\xi \alpha}{a} \left[\partial_{r}A_{a} 
- \frac{2}{3}\partial_{r}K + 6A_{a}\partial_{r}\chi  \right. \nonumber \\
  & \left. + (A_{a}-A_{b})\left(\frac{2}{r} + \frac{\partial_{r}b}{b}\right)
\right] \ , \\
	L_{3(\hat{\Delta}^{r})} &= \beta^{r}\partial_{r}\hat{\Delta}^{r}
- \hat{\Delta}^{r}\partial_{r}\beta^{r}
+ \frac{2\sigma}{3}\hat{\Delta}^{r}\hat{\nabla}_m\beta^{m} \nonumber \\
	& + 2\alpha A_{a}\hat{\Delta}^{r} - 8\pi j_{r} \frac{\xi \alpha}{a} \ ,\\
	L_{2(\Psi)} &= \partial_{r}(\alpha\Pi) \ , \\
	L_{3(\Psi)} &= \beta^{r} \partial_{r}\Psi  + \Psi\partial_{r}\beta^{r} \ , \\
	L_{2(\Pi)} &= \frac{\alpha}{ae^{4\chi}}\biggl[\partial_{r}\Psi +\Psi\biggl(\frac{2}{r}-\frac{\partial_{r}a}{2a}+\frac{\partial{r}b}{b}+2\partial_{r}\chi\biggl)\biggl]\nonumber\\
&+\frac{\Psi}{ae^{4\chi}}\partial_{r}\alpha - \alpha \biggl[\mu^{2}+\lambda\,|\Phi|^{2}+q^{2}\biggl(\frac{a_{r}^{2}}{ae^{4\chi}}\biggl)\biggl]\Phi \nonumber\\
&+2iq\alpha\biggl[\frac{a_{r}\Psi}{ae^{4\chi}}+\varphi\Pi\biggl]\,, \\
	L_{3(\Pi)} &= \beta^{r}\partial_{r}\Pi +\alpha K\Pi \ ,\\
	L_{2(\varphi)} &= -\frac{\alpha}{ae^{4\chi}}\biggl[\partial_{r}a_{r} +a_{r}\biggl(\frac{2}{r}-\frac{\partial_{r}a}{2a}+\frac{\partial{r}b}{b}+2\partial_{r}\chi\biggl)\biggl]\nonumber\\
&-\frac{a_{r}}{ae^{4\chi}}\partial_{r}\alpha, \\
	L_{3(\varphi)} &= \beta^{r}\partial_{r}\varphi +\alpha K\varphi \ , \\
	L_{2(a_{r})} &= a_{r}\partial_{r}\beta^{r}-\partial_{r}(\alpha\varphi)\ , \\
	L_{3(a_{r})} &= \beta^{r}\partial_{r}a_{r} -\alpha ae^{4\chi}\,E^{r} \ .
\end{align}
Finally, $B^r$ is evolved partially implicitly, using the updated values of 
$\hat{\Delta}^{r}$, i.e., 
$\displaystyle L_{2(B^r)} = \frac{3}{4}\partial_{t}\hat{\Delta}^{r}$ and 
$L_{3(B^r)} = 0$.

\begin{widetext}

\begin{figure}[h!]
\includegraphics[width=0.5\textwidth]{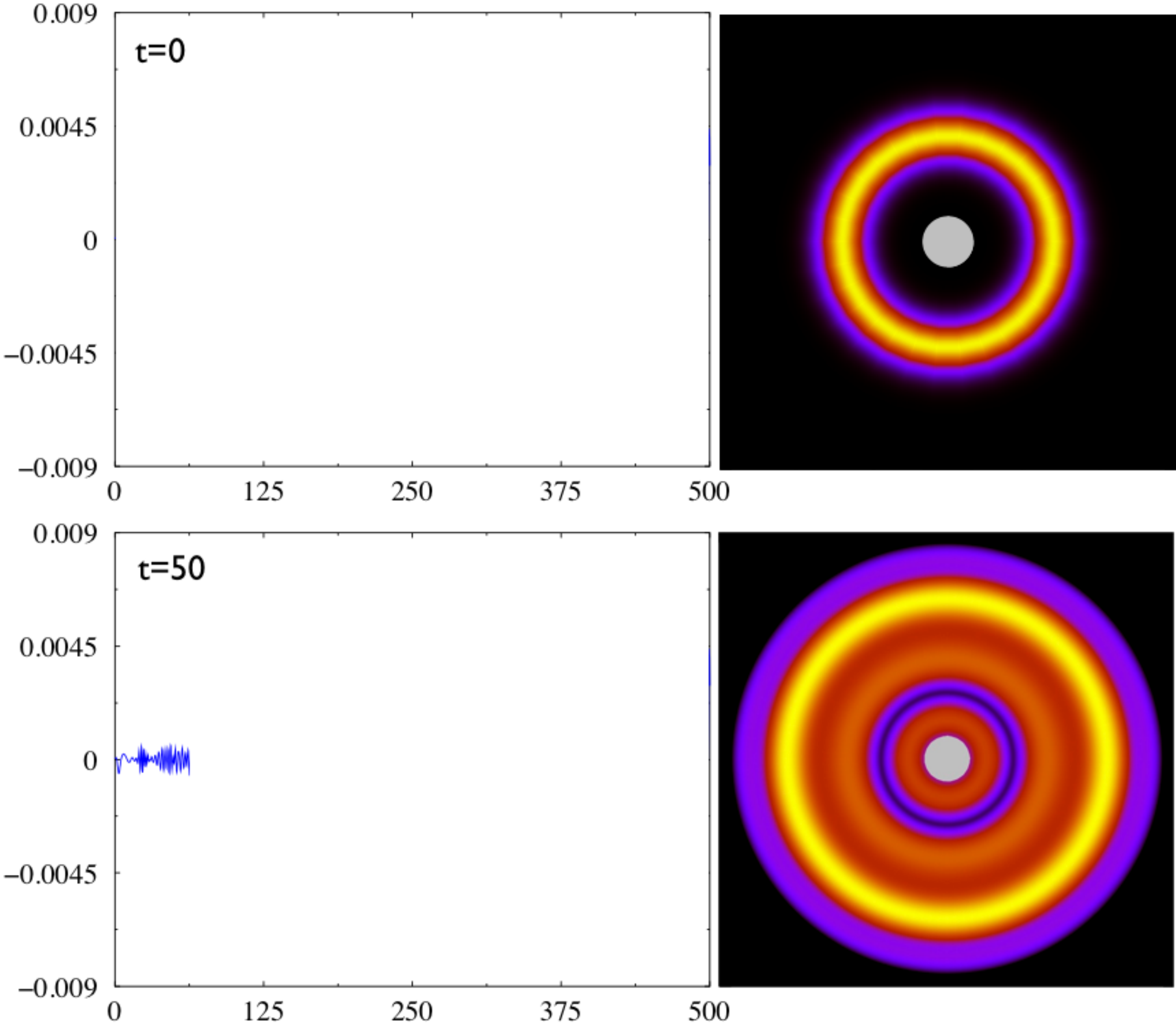} 
\includegraphics[width=0.5\textwidth]{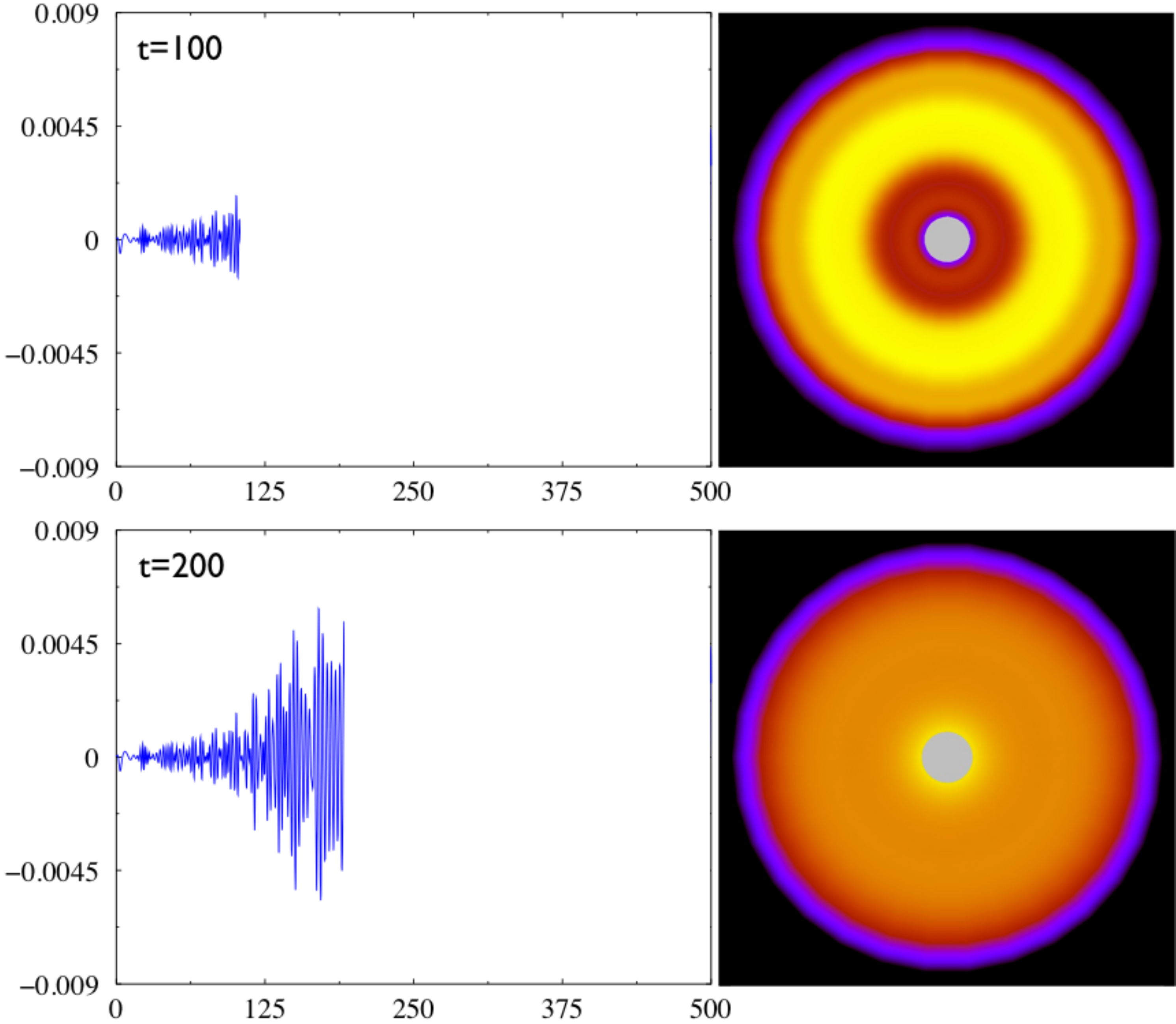} 
 \includegraphics[width=0.504\textwidth]{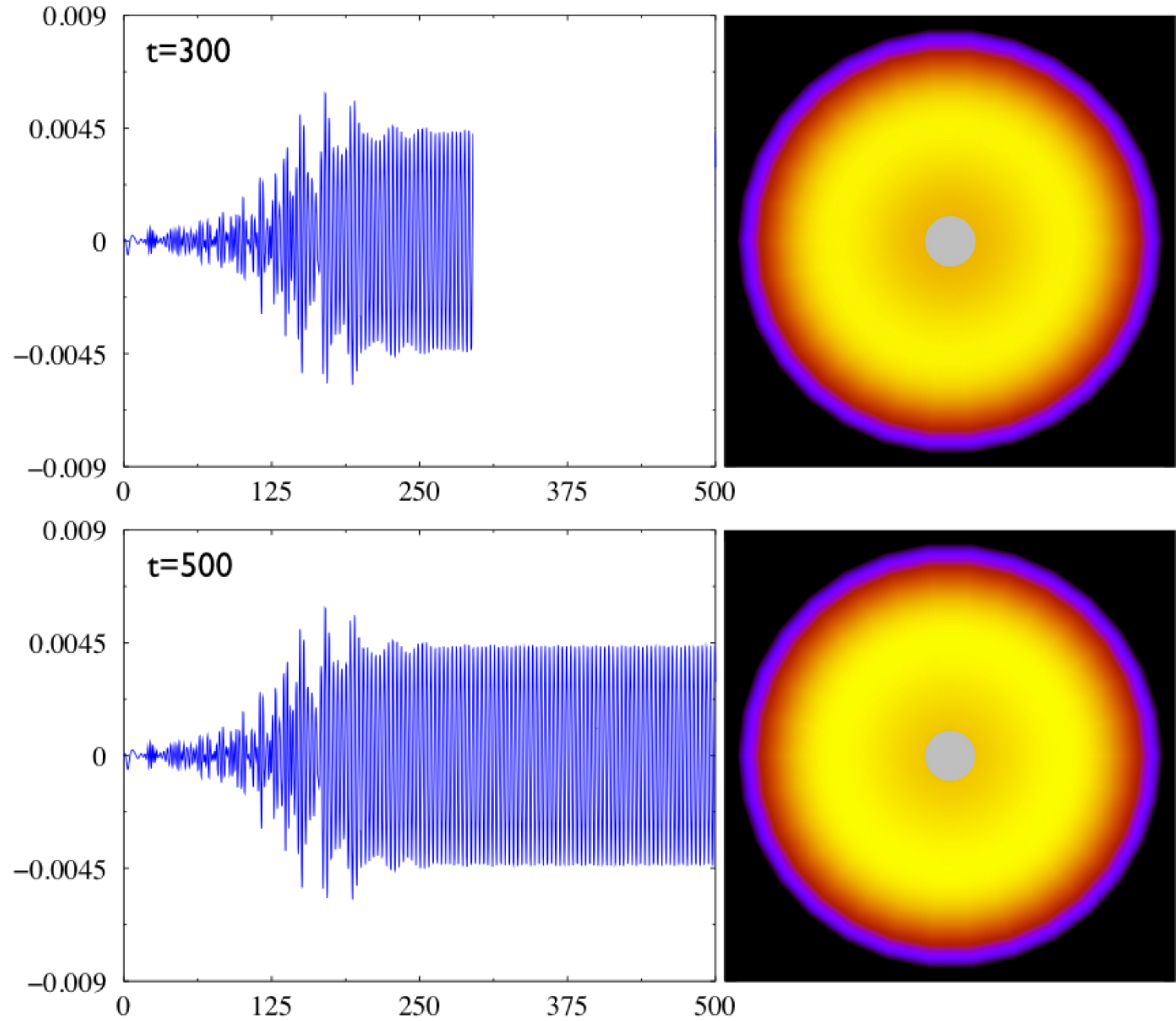}
\begin{picture} (0,0)
\put(50,255){\includegraphics[width=0.05\textwidth]{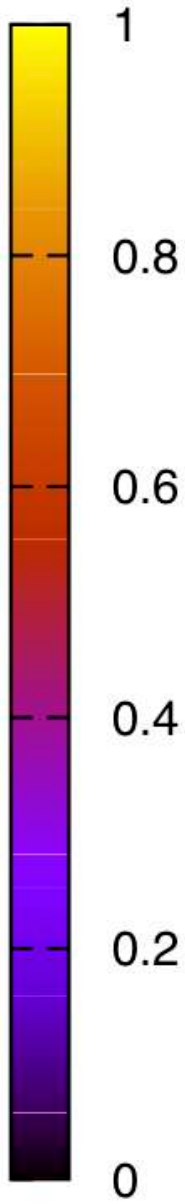} }
\end{picture}
\caption{Illustration of the  formation of a hairy BH for $qM=20$, $\lambda=0$. 
The left panels show the time series until a certain time and the right (2D) panels 
show the corresponding snapshot, at that time, of the normalized scalar field 
profile magnitude ($cf.$ colour bar). The inner white circle denotes the BH region, bounded by the apparent horizon.
}
\label{fig_evolution}
\end{figure}

\end{widetext}

\bibliography{num-rel}

\end{document}